%% file: Isomonodromic_tau_dependence.tex
\pdfoutput=1
\documentclass[10pt]{article}
\usepackage{wrapfig,cancel}
\usepackage{color}
\usepackage{graphicx}
\usepackage[colorlinks=true, pdfstartview=FitV, linkcolor=blue, citecolor=blue,urlcolor=blue]{hyperref}

\newenvironment{proof}{{\bf Proof} }{ {\bf Q.E.D.}\par\vskip 5pt}

\def\KK{\mathbf K}
\def\blue#1{\textcolor[rgb]{0,0,1}{#1}}

\usepackage{amsbsy}
\usepackage{amssymb}
\newtheorem{problem}{Problem}[section]
\def\D{\mathbb D}
\newtheorem{question}{Question}
\def\T{\mathcal T}
\newtheorem{assumption}{Assumption}[section]

\usepackage{verbatim}
\def\tr {\mathrm {Tr}}

\makeatletter
\@addtoreset{equation}{section}
\makeatother

\newtheorem{theorem}{Theorem}[section]
\newtheorem{example}{Example}[section]
\newtheorem{exercise}{Exercise}[section]

\newtheorem{lemma}{Lemma}[section]
\newtheorem{remark}{Remark}[section]

\newtheorem{proposition}{Proposition}[section]
\newtheorem{corollary}{Corollary}[section]
\newtheorem{definition}{Definition}[section]
\def\le{\left}
\def\ri{\right}
\def\ds{\displaystyle}

\def\res{\mathop{\mathrm {res}}\limits_}

\def\br{\begin{remark}}
\def\er{\end{remark}}
\def\bt{\begin{theorem}}
\def\et{\end{theorem}}
\def\bc{\begin{corollary}}
\def\ec{\end{corollary}}
\def\bx{\begin{example}\small}
\def\ex{\end{example}}
\def\bxr{\begin{exercise}\small}
\def\exr{\end{exercise}}
\def\bl{\begin{lemma}}
\def\el{\end{lemma}}
\def\dd{ \frac{{\rm d }x} {2i\pi}}
\def\bd{\begin{definition}}
\def\ed{\end{definition}}
\def\bp{\begin{proposition}}
\def\ep{\end{proposition}}
\def\be{\begin{equation}}
\def\ee{\end{equation}}
\def\&{\hspace{-15pt}&}
\def\bea{\begin{eqnarray}}
\def\eea{\end{eqnarray}}
\def\beas{\begin{eqnarray*}}
\def\eeas{\end{eqnarray*}}

\def \pa{\partial}
\def\C{{\mathbb C}}

\def\N{{\mathbb N}}

\def\wh{\widehat}

\def\Z{{\mathbb Z}}

\def\d{\,\mathrm d}
\def\K{\mathcal K}

\def\1{{\bf 1}}
\def\wt{\widetilde}

\date{}
\begin{document}
\baselineskip 15pt plus 1pt minus 1pt

\vspace{0.2cm}
\begin{center}
\begin{Large}
\fontfamily{cmss}
\fontsize{17pt}{27pt}
\selectfont
\textbf{
The dependence on the monodromy data  of the isomonodromic tau function}
\end{Large}\\
\bigskip
\begin{large} {M.
Bertola}$^{\ddagger,\sharp}$\footnote{Work supported in part by the Natural
    Sciences and Engineering Research Council of Canada
(NSERC).}\footnote{bertola@crm.umontreal.ca}.
\end{large}
\\
\bigskip
\begin{small}
$^{\ddagger}$ {\it Department of Mathematics and
Statistics, Concordia University\\ 1455 de Maisonneuve W., Montr\'eal, Qu\'ebec,
Canada H3G 1M8} \\
$^{\sharp}$ {\it Centre de recherches math\'ematiques,\  Universit\'e\ de
Montr\'eal } \\
\end{small}
\bigskip
{\bf Abstract}
\end{center}
The isomonodromic tau function defined by Jimbo-Miwa-Ueno vanishes on the Malgrange's divisor of generalized monodromy data for which a vector bundle is nontrivial, or, which is the same, a certain Riemann--Hilbert problem has no solution. In their original work, Jimbo, Miwa, Ueno provided  an algebraic construction of its derivatives with respect to isomonodromic times. However the dependence on the (generalized) monodromy data (i.e. monodromy representation and Stokes' parameters) was not derived. We fill the gap by providing a (simpler and more general) description in which all the parameters of the problem (monodromy-changing and monodromy-preserving) are dealt with at the same level.
We thus provide variational formul\ae\ for the isomonodromic tau function with respect to the (generalized) monodromy data.  
The construction applies more generally: given any (sufficiently well-behaved) family of Riemann--Hilbert problems (RHP) where the jump matrices depend arbitrarily on deformation parameters, we can construct a one-form $\Omega$ (not necessarily closed) on the deformation space (Malgrange's differential), defined off Malgrange's divisor.
We then introduce the notion of  discrete Schlesinger transformation: it means that we allow the solution of the RHP to have poles (or zeros) at prescribed point(s).  Even if $\Omega$ is not closed, its difference evaluated along the original solution and the transformed one,  is shown to be  the logarithmic differential (on the deformation space) of a function.
As a function of the position of the points of the Schlesinger transformation, yields a natural generalization of Sato formula for the Baker--Akhiezer vector  even in  the absence of a tau function, and it realizes the solution of the RHP as such BA vector.

Some exemplifications in the setting of  the Painlev\'e\ II equation and finite T\"oplitz/Hankel determinants are provided.

\vspace{0.7cm}

\vskip 15pt

\tableofcontents

\section{Introduction}
In the eighties Jimbo, Miwa and Ueno \cite{JMU1, JMU2, JMU3} derived a set of algebraic (in fact rational) nonlinear equations describing deformations of a rational connection on $\mathbb P^1$ which preserve the generalized monodromy data.
They associated to this deformation a closed differential $\omega_{_{JMU}}$ on the space of deformation parameters, namely, on the space of ''isomonodromic times'' which we denote collectively by $\vec t$. 
In the simplest case of Fuchsian singularities 
\be
\Psi'(z) = A(z)\Psi(z)\ , \ \ \Psi(\infty)=\1 \ , \qquad A(z) = \sum_{j=1}^K \frac {A_j}{z-a_j}\ ,\ \sum A_j =0
\ee
the isomonodromic deformation equations were studied by Schlesinger \cite{Schlesinger}
\be
\delta  A_k = -\sum_{j\neq k} [A_k, A_j] \frac {\delta (a_k-a_j)}{a_k-a_j} \ ,\ \ \ \delta:= \sum \d a_j \pa_{a_j} 
\label{zerocurv}
\ee 
and the Jimbo-Miwa-Ueno differential reads 
\be
\omega_{_{JMU}} = \frac 1 2 \sum_{j,k, j\neq k} \tr (A_jA_k) \frac {\delta(a_j -a_k)}{a_j-a_k}
\ee
It can be checked directly that if the matrices $A_k(\vec a)$ depend on  the position of the poles as mandated by (\ref{zerocurv}) then $\omega_{_{JMU}}$ above is a closed differential.

This differential was generalized to arbitrary (generic) rational connection in \cite{JMU1}, to which we refer the reader for details. In the above situation for Schlesinger deformations,  the locations of the poles constitute the ``isomonodromic'' parameters or times and we denote them by $\vec t$. 
This was an important achievement because of the sweeping applications of isomonodromic deformation to integrable systems (solitons solutions to KP, solutions to Toda, etc.)  Painlev\'e\ equations and, later, random matrix models. 
The (exponential) integral of this closed differential is the ``isomonodromic tau function''
\be
\tau_{_{JMU}}(\vec t; \vec m) = {\rm e}^{\int \omega_{_{JMU}}}\ . \label{swindle}
\ee
and the Painlev\'e\ property translates to the fact that $\tau_{_{JMU}}$  is a holomorphic function of the isomonodromic times that has only zeroes (away from an explicit set of times where it has  a branching behavior: in the case of (\ref{zerocurv}) this is the set of ``diagonals'' $a_j=a_k$, $j\neq k$).  In (\ref{swindle}) we have indicated that the tau function depends necessarily on the (generalized) monodromy data, denoted generically by $\vec m$; this dependence is parametric and the present paper addresses precisely the
 
\begin{question}[Naive] 
\blue{What is the dependence of $\tau_{_{JMU}}$ on the monodromy data?}
\end{question} 

The question is conceptually simple but slightly ill-posed; since what JMU really defined was only a differential (in symbolic notation)
\be
\omega_{_{JMU}} = \sum_j  f_j (\vec t; \vec m) \d t_j\ ,\ \ \pa_{t_j}f_k = \pa_{t_k} f_j
\ee
the dependence of $\tau$ on $\vec m$ is defined only up to multiplication by an arbitrary function of the $\vec m$'s only.
So a better question would be 
\begin {question}[Refined]
\label{posh}
\blue{
What is the {\em essential} dependence of $\tau_{_{JMU}}$ on the monodromy data?. Can we define an extended {\bf closed} differential $\Omega$ on the {\bf total phase space} of the problem that coincides with $\omega_{_{JMU}}$ on the isomonodromic submanifold, namely 
\bea
\omega_{ext} =  \sum_j  f_j (\vec t; \vec m) \d t_j + \sum_\nu G_\nu(\vec t, \vec m) \d m_\nu \label{extom}\\
\pa_{m_\nu} f_k = \pa_{t_k} G_\nu\ ,\ \ \pa_{m_\nu} G_\mu = \pa_{m_\nu} G_\mu.
\eea}
\end{question}
The question still admits many solutions as stated, since if we find one such extension we are still free to add any closed differential of the $\vec m$'s alone. However we may and should understand the problem in a relative setting, where the answer is taken modulo closed forms of the $\vec m$'s  which must be holomorphic on the whole space of monodromy data.
The reason for requiring holomorphicity is actually important because of the interpretation of the singularity locus of $\omega$, as we presently explain.
\paragraph{The meaning of the singularity locus of $\omega_{_{JMU}}$.}

Malgrange (for Fucshian systems) \cite{Malgrange:IsoDef1} and later Palmer (for irregular singularities)\cite{Palmer:Zeros} showed what the meaning of the zero-locus of $\tau_{_{JMU}}$ is: when $\tau_{_{JMU}}(\vec t;\vec m)=0$ then a vector bundle on $\mathbb P^1$  is nontrivial or --which is equivalent-- a Riemann--Hilbert problem is not solvable. This is the equivalent of saying that $\omega_{_{JMU}}$ has only simple poles (with ``residue'' one) away from the non-movable singularity locus (Painlev\'e\ property). The divisor where the aforementioned bundle is nontrivial is generally termed {\bf Malgrange $\Theta$ divisor} (or simply Malgrange divisor).

It is clear then that whatever extension (\ref{extom}) we are looking for, it ought to preserve the singularity locus, that is, the $G_\nu(\vec t, \vec m)$ may have singularities only where some of the $f_k$'s has, and of the same type.

So we arrive to the final formulation of a ``sensible'' problem, whose solution is the principal aim of the paper
\begin{problem}
\label{poshpr}
\blue{
Formulate a (``natural'') extended closed differential $\omega_{ext}$ (\ref{extom}) such  that its tau function (locally defined up to nonzero multiplicative constants)
\be
\tau_{ext}(\vec t, \vec m) = {\rm e}^{\int \omega_{ext}}
\ee
vanishes precisely and only on the Malgrange divisor in the extended phase space of isomonodromic times $\vec t$ and monodromy data $\vec m$.}
\end{problem}

The differential we propose is in fact very natural (see Def. \ref{defOmega} and Thm. \ref{main}): it is the pull--back of the (partial integral of the) generator of the third cohomology of the ``loop group'' to the submanifold corresponding to our total phase space. In particular it expresses the same cohomology class $c[\Theta]$   \cite{Malgrange:IsoDef1}.

We will briefly indicate some interesting problems which can be addressed and that require the knowledge of the derivative of $\tau$ with respect to non-isomonodromic times (hence the knowledge of the $G_\nu$'s). 
\paragraph{Organization of the paper.}
The heart of the paper is in fact Section \ref{RHPsec} where we introduce (recall) the definition of the Malgrange differential $\omega_{_M}$ (Def. \ref{defomega}) associated to {\em any (sufficiently well--behaved) Riemann--Hilbert problem}. Such differential is ill--defined only when the RHP is not solvable: however its exterior differential is (admits a)  smooth (extension)  over the whole ``loop group'', in particular also at the points where the RHP is not solvable \cite{Malgrange:IsoDef1}. The curvature of $\omega_{_M}$ is not zero, but it is so explicit that it is immediate to identify ``simple'' families of Riemann--Hilbert problems for which  $\omega_{_M}$ is closed. By adding to $\omega_{_M}$ an explicit smooth differential (in particular this does not change its singularity locus) we obtain a new differential $\Omega$ (Def. \ref{defOmega}) whose curvature differs slightly from that of $\omega_{_M}$ but within the same cohomology class. 

In Section \ref{SecSchles} we investigate the changes in $\Omega$ (or $\omega_{_M}$) under modification of the growth conditions for the solution of the RHP (discrete ``Schlesinger'' transformations); this allows to interpret the matrix solution of {\em any} Riemann--Hilbert problem as a Baker--Akhiezer function via a Sato--like formula, even if a notion of tau-function is not available (see Thm. \ref{tauratio} and Sect. \ref{remSato}, Sect. \ref{secHirota}).

 In the second part we specialize the setting from arbitrary Riemann--Hilbert problems to those that correspond to rational ODEs in the complex plane: Section \ref{RHPODE} is a quick reminder to the reader of the classical description of the (generalized) monodromy map, i.e.  how to associate to a rational ODE in the complex plane the set of ``Birkhoff data'' of irregular type, connection matrices, monodromy representation and Stokes' multipliers. All the material is quite standard \cite{Wasow}.

These Birkhoff data can be used viceversa (Sect. \ref{SecBirk})  to encode an ODE in a Riemann--Hilbert problem (which {\em may} or may not have a solution, although generically it does \cite{Sibuya,Wasow}).

In Section \ref{tauiso} we show that $\Omega$ is a closed differential in all the deformation parameters, which include \\
$\bullet$ Monodromy representation;
$\bullet$  Connection matrices;
$\bullet$ Stokes' matrices.

In Section \ref{MalgrangeJMU} we show that the restriction of $\Omega$ to the submanifold of isomonodromic times coincides with the differential $\omega_{_{JMU}}$ defined in \cite{JMU1}. Therefore $\Omega$ realizes the solution to our Problem \ref{poshpr}.

We conclude the paper with Section \ref{Examples}, where applications are provided to Painlev\'e\ II, (shifted) Toeplitz and Hankel determinants.

\paragraph{Acknowledgements.}
The author thanks Jacques Hurtubise and John Harnad for helpful discussion. The work was partially supported by the Natural
    Sciences and Engineering Research Council of Canada
(NSERC).

\section{Riemann--Hilbert problems}
\label{RHPsec}

A Riemann--Hilbert problem starts with the following  data. We will assume latitude in the smoothness class as this is not our primary focus.
\paragraph{The Riemann--Hilbert data}
\begin{enumerate}
\item a finite collection of smooth oriented arcs $\gamma_\nu,\ j=1\dots K$, possibly meeting at a finite number of points but always in non-tangential way (Fig. \ref{fig1} is a good example). We denote  collectively these arcs by the symbol $\Sigma \gamma$
\be
\Sigma \gamma = \bigcup_\nu \gamma_\nu
\ee
\item a collection of  $r\times r$ matrices $M_\nu(z)$, each of which holomorphic in a neighborhood of each interior point of the corresponding arc $\gamma_\nu$. We make the assumption that these matrices have  unit determinant\footnote{This is not really necessary but simplifies some matters. In some situations (typically where the determinant a rational function)  it is necessary (but simple) to relax the condition and allow any matrices in $GL_r(\C)$. We face the problem on a need-to basis.}  $\det M_\nu(z)\equiv 1$. We will denote collectively by $M(z)$ the matrix defined on $\Sigma \gamma$ that coincides with $M_\nu(z)$ on $\gamma_\nu$,  
\bea
M: \Sigma\gamma&\to& SL_r(\C)\cr
z & \mapsto & \sum_{\nu} M_\nu(z) \chi_{\gamma_\nu}(z)
\eea
where, for a set $S$,  $\chi_S$ denotes its indicator function.
\end{enumerate}
The Riemann--Hilbert problem then consists in finding a holomorphic matrix on the complement of the contours
\be
\Gamma(z):\C\setminus \Sigma\gamma\to SL_r(\C)
\ee
such that it admits (non-tangential) boundary values satisfying 
\bea
\Gamma_+(z) = \Gamma_-(z) M(z) \ \ \ z\in \Sigma\gamma\ .
\eea
The above data are insufficient to characterize uniquely the matrix $\Gamma$ (if it exists) and need to be supplemented by 
\begin{itemize}
\item growth behavior near the endpoints/intersections of the contours $\gamma_\nu$ and at $\infty$;
\item an overall normalization.
\end{itemize}
Therefore we will pose the following 
\begin{problem}[RHP]
\label{probRH}
Find a holomorphic matrix $\Gamma:\C\setminus \Sigma\gamma\to GL_n(\C)$ such that 
\begin{itemize}
\item $\Gamma_+(x) = \Gamma_-(x) M(x)$ $x\in \Sigma\gamma$;
\item $\Gamma(z)$ is uniformly bounded in $\C$;
\item $\Gamma(z_0)=\1$
\end{itemize}
\end{problem}
Typically we will choose $z_0=\infty$.

\begin{wrapfigure}{l}{0.5\textwidth}
\resizebox{0.5\textwidth}{!}{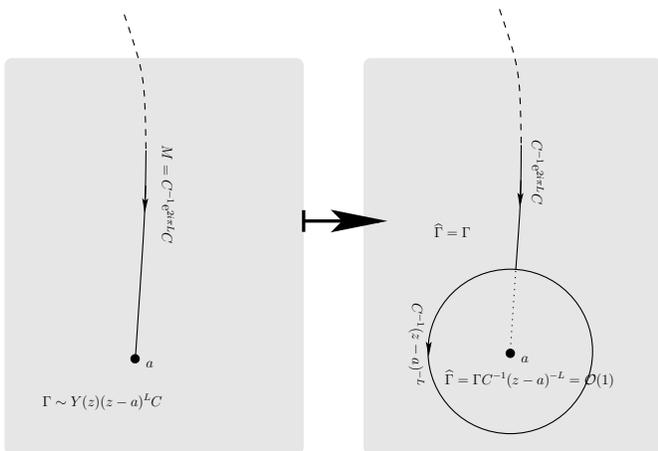}
\caption{The process of ``cloaking'' of a point of growth. The two RHPs are equivalent in the sense that one admits a solution if and only if the other does.}
\label{hidingsing}
\vskip 0cm
\end{wrapfigure}

In general there are obstructions to the solvability of Problem \ref{probRH}: for example let $c$ be a point in $\Sigma \gamma$ where several arcs $\gamma_1,\dots, \gamma_\ell$ meet (at nonzero tangential relative angles) and suppose they are oriented towards $c$. Let  $M_j(z)$, $j=1,\dots,\ell$ are the corresponding (locally analytic) jump matrices, then 
\be
M_1(c) M_2(c) \dots M_\ell(c) \neq \1\ .\label{obstrmono}
\ee
is an obstruction. This does not mean that it is impossible to find $\Gamma$ satisfying the jumps, but it cannot be bounded at $c$ in general and thus a relaxed growth constraint must be allowed at $c$.

However there are situations which are of interest for our applications where a growth behavior can be ``traded in'' for an extra contour. 
\bx
A typical example is  depicted in Fig. \ref{hidingsing}; here on the contour that terminates at $z=a$ there is a constant jump of the form $M= C^{-1} {\rm e}^{2i\pi L} C$, where $L$ is upper triangular (in Jordan form, for example). It is clear that the diagonal entries of $L$ (the {\bf exponents of formal monodromy}) are defined only up to addition of integers. This arbitrariness in fact corresponds to the {\em choice} of a growth behavior at the endpoint for the solution of the RHP. Once this choice has been made, we can always recast the problem into an equivalent one where the solution $\Gamma$ is bounded. This is achieved by adding a small circle at $a$ and re-defining $\wt \Gamma = \Gamma C^{-1} (z-a)^{-L}$ inside: in the new formulation for the problem, $\wt \Gamma$ will be bounded in a neighborhood of $z=a$ and the two problems are equivalent, in the sense that one has solution if and only if the other does, and the relation between the two solutions is also very simple.
\ex
\noindent {\bf (Loop) group structure.} Although we will not make any explicit use of the following fact, we mention that the matrices $M:\Sigma\gamma\to SL_r(\C)$ that satisfy all the condition above form a (infinite dimensional) {\bf Lie group}, akin to the usual loop group, were it not for the fact that the contour is not a (collection of) circle(s). It is convenient to introduce a symbol for this manifold (group)
\bea
\mathcal G:= \bigg\{&\& M: \bigcup \gamma_\nu\mapsto SL_r(\C),\ M\big|_{\gamma_\nu} (z) \hbox{ locally analytic}\bigg\}
\eea
Since we will only consider finite dimensional (analytic) submanifolds of $\mathcal G$ we will not dwell on the infinite--dimensional differential-geometric issues. Most notably the tangent bundle $T \mathcal G$ will be used in a rather naive form where possible issues of topological nature (arising from the infinite dimensionality) are disposed of. In fact only the tangent space to the finite dimensional submanifolds of interest will appear, and hence the point is not relevant.

\subsection{Deformations and Malgrange's form}
Suppose now that $M(z) = M(z;\vec s)$ depends holomorphically on additional parameters $\vec s$; the reader should think of this as an explicit dependence, dictated by the problem under consideration. 
The parameters $\vec s$ could be thought of as coordinates on a manifold of deformations. When -eventually- we specialize the setting these $\vec s$'s will be the isomonodromic times together with the monodromy data. On this manifold we define the one-form (differential) already used by  Malgrange \cite{Malgrange:IsoDef1}.
\bd
\label{defomega}
Let $\pa$ denote the derivative w.r.t. one of the parameters $s$. Then we define Malgrange's form $\omega_{_M}$ 
\bea
\omega_{_M}(\pa) = \omega_{_{M}}(\pa; [\Gamma]) :=  - \int_{\Sigma \gamma} \tr \bigg(\Gamma^{-1}_-(x) \Gamma_-'(x)  \Xi_\pa (x)\bigg) \dd \label{omegaM}\\
\Xi_\pa(z):= \pa M(z) M^{-1}(z)\ .\nonumber
\eea
\ed
\br
The minus sign is --of course-- conventional, but it is important because with the minus the tau-function (when it exists) has a {\em zero} and not a {\em pole} on the Malgrange divisor.
\er
\br
The definition would hold identically also for any RHP formulated on a Riemann surface.
\er
\br
The reader may frown upon the boundary value of the derivative of $\Gamma$: however, since our assumptions mandate that $M_\nu(z)$ is analytic in a neighborhood of the contour $\gamma_\nu$, it is simple to show that the derivative $\Gamma'$ also admits (bounded) boundary values.
\er
\br
In the second notation for $\omega_{_M}$ we have indicated the dependence on $\Gamma$, which depends on the choice of jump matrices  {\bf and} specifications of the growth behaviors. 
 When no ambiguity can arise, we will understand such dependence.
\er

\paragraph{Curvature of $\omega_{_{M}}$.}
The first issue is the computation of the exterior differential: the computation can be found in \cite{Malgrange:IsoDef1} but in rather abstract terms and we prefer to give a direct derivation here.
We have  first the 
\bl
\label{varGamma}
Let $\pa$ denote any vector field in the parameters of the jump matrices. Then
\be
\pa \Gamma(z) = \int_{\Sigma \gamma} \frac {\Gamma_-(x) \Xi_\pa(x) \Gamma_-^{-1}(x)}{x-z} \dd\, \Gamma(z)\label{pagamma}
\ee
and, consequently 
\be
\pa \le(\frac {\Gamma^{-1}(z)\Gamma(w)}{z-w} \ri) = \int_{\Sigma\gamma}   \frac {\Gamma^{-1}(z)\Gamma_-(x)}{z-x} \Xi_\pa(x)  \frac {\Gamma_-^{-1}(x)\Gamma(w)}{x-w}\dd\label{pakern}
\ee
\el
\begin{proof}
First of all we have 
\be
\pa \Gamma_+ = \pa \Gamma_- M + \Gamma_- \pa M\ .\label{nhjump}
\ee
This is a non-homogeneous Riemann--Hilbert problem: for convenience we take the normalization point at infinity  $\Gamma(\infty)=\1$ so that $\pa \Gamma(z) = \mathcal O(z^{-1})$. It is promptly seen that the proposed expression fulfills (\ref{nhjump}) and this last condition. The uniqueness follows from the fact that the homogeneous part  has only the trivial solution that tends to zero at $\infty$. The formula (\ref{pakern}) follows by direct application of (\ref{pagamma}) and Leibnitz rule. 
\end{proof}
\bp
\label{propsympl}
Denoting with $\pa, \wt \pa$ the derivatives w.r.t to two of the parameters $\vec s$, the exterior differential of $\omega_{_M}$ is 
\be
\eta(\pa, \wt \pa) := \pa \omega_{_M}(\wt \pa) - \wt\pa\omega_{_M}(\pa) = \frac 12 \int_{\Sigma \gamma} \tr  \le(\Xi_{\wt\pa} (x)  \frac {\d}{\d x}\Xi_{ \pa}(x) - \frac {\d}{\d x}\Xi_{\wt\pa} (x)  \Xi_{\pa}(x) \ri)\dd
\label{sympl} 
\ee
\ep
\begin{proof}
Let $\pa,\wt \pa$ be two commuting vector fields so that 
\be
\delta \omega (\pa, \wt \pa) = \pa \omega (\wt \pa) - \wt \pa \omega(\pa).
\ee
We have 
\bea
\pa\omega_{_M} (\wt \pa) &\&= 
-\pa \int_{\Sigma\gamma} \tr \le( \Gamma^{-1}_-(x) \Gamma'_- (x)\Xi_{ \wt \pa}(x)\ri)\dd = \\
=&\&
-\int_{\Sigma\gamma} \tr \le( \blue{\pa \le(\Gamma^{-1}_-(x) \Gamma'_-(x) \ri)} \Xi_{\wt \pa}(x)\ri)\dd + \tr \le( \Gamma^{-1}_-(x) \Gamma'_- (x) \blue{\pa\le(\Xi_{ \wt \pa}(x)\ri)}\ri)\dd = \\
=&\& -\int_{\Sigma\gamma} \tr \le( 
\int_{\Sigma\gamma} \frac{\K(x,y) \Xi_\pa(y)\K(y,x)\Xi_{\wt\pa}(x)}{(x-y)^2} \d y
\ri)\bigg|_{x=x_\pm}\dd 
-
\int_{\Sigma\gamma}  \tr \le( \Gamma^{-1}_-(x) \Gamma'_-(x) \le( \wt
\pa \Xi_\pa(x)  +  \le[\Xi_{\pa},\Xi_{\wt \pa}\ri]\ri)\ri)\dd \label{secondder}
\eea
where we have introduced the convenience notation 
\be
\K(x,y) = \Gamma_-^{-1}(x) \Gamma_-(y)\ .
\ee
Continuing our computation we find
\bea
\pa \omega_{_M} (\wt \pa) - \wt \pa \omega_{_M} (\pa) =&\& 
  \int_{\Sigma\gamma}\int_{\Sigma\gamma}  \tr \le( 
\frac{ \K(x,y)\Xi_{ \wt \pa} (y)\K(y,x)\Xi_{ \pa}(x)}{(x-y)^2} \d y\,\ri) \bigg|_{x=x_-}
 \dd+
\cr
 &\&
-
\int_{\Sigma\gamma} 
\int_{\Sigma\gamma} \tr \le( \frac{ \K(x,y) \Xi_\pa(y)\K(y,x)\Xi_{\wt \pa}(x)}{(x-y)^2}\d y \ri) \bigg|_{x=x_-}
\dd +\cr
&&+ \int_{\Sigma\gamma}  \tr \le( \Gamma^{-1}_- \Gamma'_ - \bigg[ \Xi_{\wt \pa} ,\Xi_{\pa}\bigg]\ri)\dd \label{sasda}
\eea
In the first two terms the order of integration is important since the kernel is singular due to the denominator $(x-y)^2$. Note that --by our assumptions-- the jump matrices have analytic continuation in a neighborhood of the contours. A standard computation involving residues yields the expression 
\bea
&\& \int_{\Sigma\gamma}\int_{\Sigma\gamma} \tr \le( 
\frac{ \K(x,y)\Xi_{ \wt \pa} (y)\K(y,x)\Xi_{ \pa}(x)}{(x-y)^2}\bigg|_{x=x_-} 
 \,\ri) \wh {\d y} \dd - 
 \int_{\Sigma\gamma} 
\int_{\Sigma\gamma} \tr \le( \frac{ \K(x,y) \Xi_\pa(y)\K(y,x)\Xi_{\wt \pa}(x)}{(x-y)^2}\bigg|_{x=x_-}
\ri) \wt {\d y} \,\dd =
 \cr
 &\&  = 
  \int_{\Sigma\gamma}  \frac {\d}{\d y}\tr  \bigg(\K(x,y) \Xi_\pa(y)\K(y,x)\Xi_{\wt \pa}(x)\bigg)\bigg|_{y=x}\dd - \frac 1 2 \tr  \bigg( \Xi_\pa(x)\Xi_{\wt \pa}(x)\bigg)\bigg|_{x\in \delta\Sigma\gamma}=\label{bt}\\
 &\&=   \int_{\Sigma\gamma}\tr \le[\Gamma_-^{-1}(x)\Gamma_-'(x) \Xi_\pa(x) \Xi_{\wt \pa}(x) - \Xi_\pa(x) \Gamma_-^{-1}(x) \Gamma_-'(x) \Xi_{\wt \pa}(x)  + \Xi_{\pa}' (x)  \Xi_{\wt \pa}(x)\ri]\dd
 -
  \frac 1 2 \tr  \bigg( \Xi_\pa(x)\Xi_{\wt \pa}(x)\bigg)\bigg|_{x\in \delta\Sigma\gamma}=\cr
 &\& = 
  \int_{\Sigma\gamma}\tr \le(\Gamma_-^{-1}(x)\Gamma_-'(x)\le[ \Xi_\pa(x), \Xi_{\wt \pa}(x)\ri]\ri) \dd 
+  \int_{\Sigma\gamma}\tr\le(  \Xi_{\pa}' (x)  \Xi_{\wt \pa}(x)\ri)\dd   
-   \frac 1 2 \tr  \bigg( \Xi_\pa(x)\Xi_{\wt \pa}(x)\bigg)\bigg|_{x\in \delta\Sigma\gamma} = \\
 &\& =    \int_{\Sigma\gamma}\tr \bigg(\Gamma_-^{-1}(x)\Gamma_-'(x)\le[ \Xi_\pa(x), \Xi_{\wt \pa}(x)\ri]\bigg) \dd  
+  \frac 1 2 \int_{\Sigma\gamma}\tr\le(  \Xi_{\pa}' (x)  \Xi_{\wt \pa}(x) - \Xi_{\pa} (x)  \Xi_{\wt \pa}'(x) \ri)\dd  
 \eea
 The notation for the last term in (\ref{bt}) is the evaluation of the ``boundary term'' at the endpoints of all the $\gamma_\nu$ according to the rules of integration by parts (i.e. with the appropriate signs depending on the orientations of the contours).
 
Thus, resuming (\ref{sasda}), we have
\bea
\eta(\pa, \wt \pa):= \pa \omega_{_M} (\wt \pa) - \wt \pa \omega_{_M} (\pa) =&\&   \frac 1 2 \int_{\Sigma\gamma}\tr\le(  \Xi_{\wt\pa} (x)  \frac {\d}{\d x}\Xi_{ \pa}(x) - \frac {\d}{\d x}\Xi_{\wt\pa} (x)  \Xi_{ \pa}(x) \ri)\dd  
\eea
\end{proof}

\paragraph{The Theta divisor.} The reader should observe and compare the expressions (\ref{sympl}) and (\ref{omegaM}): the crucial point is that $\omega_M$ (\ref{omegaM}) is not defined whenever the RHP does not admit a solution. On the contrary, as a sort of ``miracle'' the expression (\ref{sympl}) for its curvature is defined and holomorphic for any group--valued matrix  $M(z)$, whether or not the RHP is solvable. In other words, $\eta$ is a closed two--form which is smooth also on the Malgrange divisor, whereas $\omega_M$ is undefined at those points. 
Malgrange proved \cite{Malgrange:IsoDef1} (and Palmer generalized \cite{Palmer:Zeros}) that $\omega_M$ has a ``pole'' on such divisor: evidently, in the curvature such a pole disappears.

\paragraph{Tau function}
It is also apparent that $\omega_M$ is in general not a closed one-form on the whole (infinite--dimensional) manifold $\mathcal G$: however, it may become closed when restricted to suitable submanifolds $\mathcal L\hookrightarrow \mathcal G$. 
Since the curvature $\eta$ (\ref{sympl}) is explicitly computable, it is easy to verify for a given explicit submanifold $\mathcal L$ whether $\omega_M\big|_{\mathcal L}$ is closed or not. 
When such restriction turns out to be closed we can define  (locally) a function on $\mathcal L$ by 
\be
\tau_{\mathcal L} = {\rm e}^{\int \omega_{_M}\big|_{\mathcal L}}
\ee
By Malgrange's (Palmer's) results, this function (defined up to nonzero multiplicative constant) will vanish at the intersection with the $\Theta$ divisor 
\be
\tau_{_{\mathcal L}}(\vec s)=0 \ \Leftrightarrow \ \ \vec s \in \mathcal L \cap \Theta
\ee

Suppose, however that $\omega_{_M}\big|_{\mathcal L}$ is not closed; we may still seek another differential $\theta$ that ``cures'' the curvature
\be
\d \le(\omega_{_M} + \vartheta \ri)\big|_{\mathcal L} \equiv 0
\ee
and then proceed with the construction of the tau function as before. 

Ideally such differential should be smooth on the whole $\mathcal G$ so that it does not change the cohomology class of $\omega_M$ and also does not change the singularity structure; this way the tau function will still vanish only at $\mathcal L \cap \Theta$ and our goal is met.

Although it is not a pedagogical approach, since we know what $\mathcal L$ will be for our purposes and we have already found $\vartheta$, we will describe it directly here.

\paragraph{Changing the curvature of $\omega_{_M}$.}
Consider the one--form $\vartheta$ that, evaluated on a vector $\pa$ yields 
\be
\vartheta(\pa):=- \frac 1 2 \int_{\Sigma\gamma} \tr \le(M'M^{-1} \pa M M^{-1}\ri)\dd\label{recurv}
\ee
Its curvature is the two--form
\be
\delta \theta(\pa,\wt \pa)  = \pa \theta(\wt \pa) - \wt\pa\theta(\pa) =\frac 12  \int_{\Sigma\gamma} \tr \le(\wt\pa M'  M^{-1}   \pa M M^{-1} -  \pa M' M^{-1} \wt\pa M M^{-1} \ri)\dd
\ee
which the reader can verify using Leibnitz rule; here the prime denotes (as always) the derivative with respect to $z$. The important (but trivial) additional observation  is that $\vartheta$ is a smooth differential on the whole $\mathcal G$, since it does not require the solution of a RHP.
\bd
\label{defOmega}
The modified Malgrange differential is defined as $\Omega:= \omega_{_M} +\vartheta$.
\ed
\br
A direct computation shows that 
\be
\int_{\Sigma \gamma} \tr \le( \Gamma_-^{-1} \Gamma_-' \pa M M^{-1}  \ri) \dd - \int_{\Sigma \gamma} \tr \le(\Gamma_+^{-1} \Gamma_+' M^{-1} \pa M  \ri) \dd=   -\int_{\Sigma\gamma}
\tr \le(M' M^{-1} \pa M M^{-1}\ri)\dd = 2\vartheta(\pa)
\ee
so that $\Omega$ can be written in the more symmetric form  
\be
\Omega(\pa;[\Gamma]) = -\frac 1 2 \int_{\Sigma\gamma} \tr \le( 
 \Gamma_-^{-1} \Gamma_-' \pa M M^{-1} +\Gamma_+^{-1} \Gamma_+' M^{-1} \pa M 
\ri)\dd
\ee
\er
Combining the computation of the curvature of $\omega_{_M}$ (Prop. \ref{propsympl}) with the  curvature of $\vartheta$ above we have the simple
\bp
\label{propcurv}
The curvature of the modified Malgrange form is 
\be
\delta \Omega(\pa,\wt \pa) = \pa \Omega(\wt \pa)  - \wt \pa \Omega( \pa)  = \frac 1 2 \int_{\Sigma \gamma} \tr \bigg(M' M^{-1} \le[\pa M M^{-1} , \wt  \pa M M^{-1}\ri]\bigg)\dd
\ee
\ep
\br
Referring to \cite{Malgrange:IsoDef1} Thm. 5.5
the curvature $\delta \Omega$  clearly lies in the same cohomology class as $\delta \omega_{_M}$ since the two differ by a smooth exact differential. If we define $\gamma$ as the total Maurer--Cartan form $
\gamma  = M'M^{-1} \d x + \sum \pa_j M M^{-1}\d s_j$ 
one has the expression 
\be
\delta \Omega = \frac 1{12\pi i} \int_{\Sigma\gamma} \tr (\gamma\wedge \gamma \wedge \gamma)
\ee
\er

\subsection{Submanifolds of $\mathcal G$ where $\Omega$ is closed}

Looking at the formula of the curvature form of Prop. \ref{propcurv}  a nontrivial class consists  of $M(z)$ that on each arc of $\Sigma \gamma$ reduce to one of the following forms:
\begin{itemize}
\item ({\bf Piecewise triangular}) Matrices of the forms
\be 
M_\nu (z) =  P_\nu(\1 + N_\nu (z))P_\nu^{-1}\ ,\ \ z\in \gamma_\nu
\ee
where $N_\nu(z)$ are upper--triangular analytic matrices and $P_\nu$ is any {\bf constant} permutation matrix (i.e. an element of the Weyl group for $SL_r$)
\item ({\bf Constants}) matrices independent of $z$;
\item ({\bf Torals}):  matrices $M_\nu = D_\nu(z)$ with $D_\nu(z)$ diagonal matrices or any conjugation thereof by an arbitrarily chosen but fixed matrix.
\end{itemize}

We see in Section \ref{RHPODE} that any (generic) rational ODE can be encoded in a Riemann--Hilbert problem with jumps of the form indicated here above. Therefore $\Omega$ on these submanifolds yields a closed differential.

We will also show that its restriction to the {\em isomonodromic} submanifolds coincides with $\omega_{_{JMU}}$.
\subsection{``Schlesinger'' transformations}
\label{SecSchles}
The aim of this section is to compare the differential $\omega_{_M}$ on two RHP for $\Gamma, \wt \Gamma$ defined by the {\bf same jumps} but {\bf different growth behaviors}, in the form of integer powers for columns at various points. We note immediately that two such solution differ (if both existent) by a {\bf left multiplication} by a rational matrix $R(z)$; indeed --having $\Gamma, \wt \Gamma$ the same jumps by assumption-- the matrix 
\be
R(z) := \wt \Gamma(z)\Gamma^{-1}(z) 
\ee
is an analytic function in $\C$ taken away the points where $\wt \Gamma$ has different growth. If this difference in growth is polynomial (as we seek now) then $R(z)$ is forced to be rational.

We now make the following observation: let $R(z)$ be a rational matrix  such that the divisor of all poles of  $R, R^{-1}$ consists of the point $c_1,\dots, c_K\not \in \Sigma \gamma$.   Define 
\be
\wt \Gamma(z):= R(z) \Gamma(z)
\ee
Quite clearly $\wt \Gamma$ solve a different RHP with the same jumps but different growth at the poles of $R$.
The difference between the $\omega_{_{M}}$ evaluated along the two different solutions is given by 
\bea
{\omega}_{_{M}}(\pa;[\wt \Gamma]) - \omega_{_{M}}(\pa;[\Gamma]) =
- \int_{\Sigma\gamma} \tr \le( \wt  \Gamma_-^{-1} \wt \Gamma_-' \pa M M^{-1}\ri)
 +  \int_{\Sigma\gamma} \tr \le( \Gamma_-^{-1} \Gamma_-' \pa M M^{-1} \ri) =\cr
=
- \int_{\Sigma\gamma} \tr \le( \Gamma_-^{-1} R^{-1}R' \Gamma_- \pa M M^{-1}\ri) =
-  \int_{\Sigma\gamma} \tr \le( R^{-1}R' \Gamma_- \pa M M^{-1} \Gamma_-^{-1}\ri) =\cr
= \int_{\Sigma\gamma} \tr \le( R^{-1}R'\le(\pa \Gamma_+ \Gamma_+^{-1} -  \pa \Gamma_- \Gamma_-\ri) \ri) = \sum_{j=1}^{K} \res{z=c_j} \tr \le(R^{-1} R' \pa \Gamma \Gamma^{-1}\ri)\label{deltaomega}
\eea
Note that -since $\Omega$ (Def. \ref{defOmega}) differ by $\omega_{_M}$ only in an explicit term $\vartheta$ that depends {\em only} on the jump matrices, we have 
\be
{\omega}_{_{M}}(\pa;[\wt \Gamma]) - \omega_{_{M}}(\pa;[\Gamma]) =\Omega(\pa;[\wt \Gamma]) - \Omega(\pa;[\Gamma]) 
\ee 
We now proceed with the definitions in the title of the section and specialize the class of rational matrices $R(z)$.
\bd
Given two distinct points $\xi\neq \eta$ and two (possibly formal) series 
\bea
Y_\xi(z) = G_\xi\le(\1 + \sum_{\ell=1}^{\infty} Y_{\xi;\ell} {z_\xi}^\ell\ri)\ ,\qquad 
Y_\eta(z) = G_\eta \le(\1 + \sum_{\ell=1}^{\infty} Y_{\eta;\ell} {z_\eta}^\ell\ri)\\
z_x:= (z-x),\ \ z_{\infty} := \frac 1 z
\eea
an elementary Schlesinger transformations at two distinct  points $\xi\neq \eta$ is a rational matrix $R(z)$ such that  
\be
R(z) Y_\xi(z) = \wh Y_\xi(z) z_\xi^{E_i}\ ,\qquad R(z)Y_\eta(z) = \wh Y_\eta(z) z_\eta^{-E_j} 
\ee
where $\wh Y_\bullet$ denote formal series of the same form. In the case of $\infty$ we have $G_\infty= \1=\wh G_\infty$. If neither $\xi\neq \infty \neq\eta$ then we impose also $R(\infty)=1$.
\ed
[For the case $\xi=\eta$ the definition should be modified in an obvious way: please see below].

The problem is  purely of algebraic nature, and not a very difficult one: the computation is contained in [\cite{JMU2}, App. A] and the derivation will not be reported (we will give below the relevant results).

Suppose now we have a RHP for $\Gamma$ with jump matrices $M$ as in Prob. \ref{probRH}: at points $\xi \neq \Sigma\gamma$ the solution $\Gamma$ to Prob. \ref{probRH} yields a (convergent) power series which can be used as input for the above procedures.

If $\xi \in \Sigma\gamma$ we cannot have a (even formal) series since --in general-- not even the {\em value} of $\Gamma$ is well defined at $z=\xi$. 
\begin{wrapfigure}{r}{0.3\textwidth}
\resizebox{0.3\textwidth}{!}{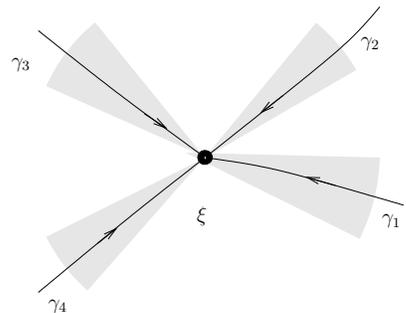}
\caption{Illustration of the sectors of analyticity for $M_j$ where the decay (\ref{decay}) should be valid.}
\label{goodpointfig}
\end{wrapfigure}
In the applications to ODEs the following  situation occurs; let $\xi\in \Sigma \gamma $ be a point where  $\gamma_1,\dots, \gamma_\ell$ meet ($\ell\geq 1$). Suppose that the jump matrices are analytic in small sectors centered at $\xi$ containing the direction of approach  and that --in said sector-- (see Fig. \ref{goodpointfig})
\be
M_j(z) = \1 + \mathcal O( (z-\xi)^{\infty})\ .\label{decay}
\ee
 Then it is not hard to see\footnote{By moving slightly the jump --which can be done due to the local  analyticity of the jump matrix-- one sees that the solution $\Gamma$ can be ``continued'' analytically across the jump from the left and from the right. The ratio of these two extension in the common sector of analyticity differs from the identity by exponentially small terms, which are transparent to any asymptotic expansion.} that the solution $\Gamma(z)$ has the {\bf same} asymptotic expansion in each of the sectors at $\xi$ separated by the incoming arcs 
\be
\Gamma(z) \sim  G_\xi\le(\1 + \sum_1^\infty Y_{\xi;j} (z-\xi)^j\ri) =: \wh Y_\xi(z)\label{goodpoint}.
\ee
We will allow to perform Schlesinger transformations involving either points $\xi$ not in $\Sigma\gamma$ or points where the condition (\ref{decay}) is met, so that the solution $\Gamma$ defines unambiguously a (formal) analytic series centered at the point. 
\bd
\label{eleSchles}
The elementary Schlesinger transformation $\le\{ \xi\ \eta\atop i\ j \ri\} $ for the solution $\Gamma$ of Problem \ref{probRH} is the solution (if it exists) of the following new RHP (where $z_\xi = (z-\xi)$ if $\xi\neq \infty$ and $z_\infty:= \frac 1 z$)

\bea
\begin{array}{c|c}
\xi\neq \eta & \xi=\eta\ \ (i\neq j)\\
\hline\\
\begin{array}{l}
\ds \wt \Gamma _+  = \wt \Gamma_-M \\[4pt]
\ds  \wt \Gamma(z) = \mathcal O(1) {z_\xi}^{-E_i}\ ,\ \ z\sim \xi\\[4pt]
 \wt \Gamma(z) = \mathcal O(1) {z_\eta}^{E_j}\ ,\ \ z\sim \eta
 \end{array}
 &
\begin{array}{l}
 \wt \Gamma _+  = \wt \Gamma_-M \\[4pt]
 \wt \Gamma(z) = \mathcal O(1) {z_\xi}^{-E_i+E_j}\ ,\ \ z\sim \xi
 \end{array}
 \end{array}
\eea
where the normalization is fixed by requiring that $\Gamma(z) \sim \1 + \mathcal O(z^{-1})$ if $\xi\neq \infty\neq \eta$ or --if either $\xi $ or $\eta$ are infinity-- that the term indicated by $\mathcal O(1)$ above is actually $\1 + \mathcal O(z^{-1})$.
\ed
Here and below we use the notation $E_{ij}$ for the elementary matrices (with a $1$ on the $i$-th row, $j$-th column) and $E_i:= E_{ii}$.
It is immediately seen that 
\be
\wt \Gamma(z) = R(z) \Gamma(z)
\ee
with $R(z)$ rational: indeed the ratio $\wt \Gamma(z) \Gamma^{-1}(z)$ for the transform $\le\{\xi\ \eta\atop i\ j \ri\}$ does not have jumps and may have at most a simple pole at $z=\xi$. The matrix $R(z)$ can be computed directly from the (possibly formal) series-expansion of $\Gamma$ at $\xi,\eta$ (see \cite{JMU2}. We give below a more compact formula that the reader can check autonomously
\bp[Cf. App. A, B in \cite{JMU2}]
\label{elemSchles}
The left-multiplier matrix $R(z)$ that implements the elementary Schlesinger transform $\le\{ \xi\ \eta\atop i\ j \ri\}$ is given by 
\bea
\begin{array}{c|c}
\le\{ \xi\ \eta\atop i\ j \ri\} \ \ \ (\xi\neq \eta) & \le\{ \xi\ \xi\atop i\ j \ri\}\ \ (i\neq j)\\
\hline\\
\ds R(z) = \1 + \frac {\xi -\eta}{(\Gamma^{-1}(\xi)\Gamma(\eta))_{ij}} \frac {\Gamma(\eta) E_{ji} \Gamma^{-1}(\xi)} {z-\xi}
 &
\ds R(z) = \1 + \frac 1{(\Gamma^{-1}(\xi)\Gamma'(\xi))_{ij}}  \frac {\Gamma(\xi) E_{ji} \Gamma^{-1}(\xi)} {z-\xi}\\[16pt]
\ds R^{-1}(z) = \1 -  \frac {\xi -\eta}{(\Gamma^{-1}(\xi)\Gamma(\eta))_{ij}} \frac {\Gamma(\eta) E_{ji} \Gamma^{-1}(\xi)} {z-\eta} &
\ds R^{-1}(z) = \1 - \frac 1{(\Gamma^{-1}(\xi)\Gamma'(\xi))_{ij}}  \frac {\Gamma(\xi) E_{ji} \Gamma^{-1}(\xi)} {z-\xi}\\[15pt]
\ds \det R(z) = \frac {z-\eta}{z-\xi} & \det R = 1
\end{array}
\eea
\bea
\begin{array}{c|c}
\le\{ \xi\ \infty\atop i\ j \ri\} \ \ \ (\xi\neq \infty)  &\le\{  \infty\ \eta \atop i\ j \ri\} \ \ \ (\eta\neq \infty) \\
\hline\\
\ds R(z) = \1 - E_{jj} + \frac {R_1}{z-\xi} 
 &
\ds R(z)  = E_{ii} (z-\eta) + R_0\\
 R_1 = \frac{-1}{(\Gamma(\xi)^{-1})_{ij}}
\le[(\1 - E_{jj})\Gamma'(\infty)-\1 \ri]E_{ji}\Gamma^{-1}(\xi)
&
\ds R_0 =\le(\1 - E_{ii} \Gamma'(\infty)\ri)\le(\1 - \frac{\Gamma(\eta)E_{ji}}{\Gamma_{ij}(\eta)}\ri)\\[20pt]
R^{-1}(z) = E_{jj} (z-\xi)  + R_0 & R^{-1}(z) = \1 - E_{ii} + \frac {R_1}{z-\eta}\\[10pt]
R_0 =\le(\1  - \frac {E_{ji} \Gamma^{-1} (\xi)}{\Gamma^{-1}_{ij}(\xi)}\ri)\le(\1 + \Gamma'(\infty) E_{jj}\ri)
  & R_1 = \frac 1{\Gamma(\eta)_{ij}} \Gamma(\eta) E_{ji}\le[\Gamma'(\infty) (\1 -E_{ii})+ \1 \ri]  \\[10pt]
\ds \det R(z) = \frac1{ z-\xi}  & \det R(z) = (z-\eta)
\end{array}
\eea
where we have denoted by $\Gamma'(\infty)$ the derivative in the local parameter, that is
\be
\Gamma(z) = \1 + \frac 1 z \Gamma'(\infty) + \dots\ .
\ee
(the formul\ae\ for $\le \{\infty\ \infty\atop i \ j\ri\}$ can be found in loc cit.)
\ep

If $\xi$ (or $\eta$ or both) belong to $\Sigma\gamma$ where condition (\ref{decay}) is in place, the computation leading to (\ref{deltaomega}) can be still carried out with minor modifications in the process but not in the result.
First of all note that the integral $\omega_{_M}(\pa, [\wt \Gamma])$ is still convergent since $\Xi_\pa = \pa M M^{-1} = \mathcal O( (x-c)^{\infty})$  (recall that we assumed $M = \1 + \mathcal O ((x-c)^\infty)$) and hence the additional algebraic growth of $\wt \Gamma^{-1} \wt\Gamma'$ along the contours incident at $c$ is still integrable when multiplied by $\Xi_\pa$.

Let $\Sigma\gamma_\epsilon$ denote the contours that lie outside of $\epsilon$ disks centered at the points $z=\xi,\eta$ (possibly the same)  of the Schlesinger transform; we then have 
\bea
\omega_{_{M}}(\pa;[\wt \Gamma]) - \omega_{_{M}}(\pa ;[\Gamma]) = -\lim_{\epsilon\to 0} \int_{\Sigma\gamma_\epsilon} \tr \le(\le(\wt \Gamma_-^{-1}\wt  \Gamma_-' - \Gamma_-^{-1} \Gamma_-'\ri)\Xi_\pa \ri)\dd = 
\cr
 = -\lim_{\epsilon\to 0} \int_{\Sigma\gamma_\epsilon} \tr \le( \Gamma_-^{-1} R^{-1}R' \Gamma_- \pa M M^{-1}\ri) =
 -  \lim_{\epsilon\to 0} \int_{\Sigma\gamma_\epsilon}\tr \le( R^{-1}R' \Gamma_- \pa M M^{-1} \Gamma_-^{-1}\ri) =\cr
= \lim_{\epsilon\to 0} \int_{\Sigma\gamma_\epsilon} \tr \le( R^{-1}R'\le(\pa \Gamma_+ \Gamma_+^{-1} - \pa \Gamma_- \Gamma_-\ri) \ri) =   \lim_{\epsilon\to 0} \oint_{|z-\xi|=|z-\eta|=\epsilon} \tr \le(R^{-1} R' \pa \Gamma \Gamma^{-1}\ri) \label{Schlessw}
\eea
In the computation of the limit (\ref{Schlessw}) we can replace $\pa \Gamma \Gamma^{-1}$ by a suitable truncation of the asymptotic series at $z=\xi,\eta$ (which do not depend on the direction of approach under our assumptions for $M$), committing an $o(1)$ error as $\epsilon \to 0$. 
Then the limit equals the formal residue
\be
\omega_{_{M}}(\pa;[\wt \Gamma]) - \omega_{_{M}}(\pa ;[\Gamma]) =\res{z=\xi} \tr\le(R^{-1}R' \pa  \wh Y_\xi(z)\wh Y_\xi(z))^{-1} \ri)\d z  +  \res{z=\eta} \tr\le(R^{-1}R' \pa  \wh Y_\eta(z)\wh Y_\eta(z))^{-1} \ri)\d z
\ee
Here the residue means simply the coefficient of the power $-1$ of the local parameter (note that $R^{-1} R'$ has at most a double pole and hence the formal residue involves at most the first two terms in the formal series $\wh Y$). 
In case only one point is involved in the Schlesinger transformation we have simply only one residue at the end.

\br
In the case of a Schlesinger transformation involving two distinct points the determinant of the solution $\det \Gamma$ cannot remain constant, since $R(z)$ has non-constant determinant. This does not pose any significant problem as we explain presently.

The modification can be explained as follows: to account for the different power-law of the columns at the two points $z=\xi, z=\eta$, small counterclockwise circles around those points should be added to $\Sigma\gamma$  imposing additional jumps of the form 
\be
M_1(z) = (z-\xi)^{-E_i}\ ,\ \ |z-\xi|=\epsilon\ ,\ \ M_2(z) = (z-\eta)^{E_j}\ ,\ \ |z-\eta|=\epsilon\ ,
\ee 
conjugating by the same matrices any jump of a contour that passes within said circles.
Of course these jumps do not have unit determinant and hence the {\em uniqueness} of the solution must be argued in a different way from the one used in the unimodular case.
However the modification in the reasoning is only minor: for a solution $\Gamma(z)$ of the new RHP we see that $\det \Gamma(z)$ is analytic and bounded everywhere, except for jumps on the new circles where
\be
\det \Gamma_+(z) = \det \Gamma_-(z) (z-\xi)^{-1} \ |z-\xi|=\epsilon\ \hbox{and}\ \ \det \Gamma_+(z) = \det \Gamma_-(z) (z-\eta)\ ,\ \ |z-\eta|=\epsilon
\ee
This means that $\det \Gamma(z)$ admits analytic continuation in the interior of the two disks, with a simple pole at $z=\eta$ and a simple zero at $z=\xi$, plus the condition $\det \Gamma(\infty)=1$. This forces $\det \Gamma(z) \equiv \frac {z-\xi}{z-\eta}$ outside of the disks, $\det \Gamma(z) = (z-\xi)$ for $|z-\eta|<\epsilon$ and viceversa $\det \Gamma(z) = \frac 1{z-\eta}$ for $|z-\xi|<\epsilon$. Any solution will have the same determinant and hence the uniqueness is established along the same way used previously.
\er

A direct computation based on (\ref{deltaomega}) yields the following theorem, which is simply a rephrasing of an homologous theorem in \cite{JMU2}, with the proper extension of understanding to the setting of RHPs.

\bt[Thm. 4.1 in \cite{JMU2}]
\label{tauratio}
Given two  RHPs  related by the elementary Schlesinger transformation $\le\{\xi\ \eta\atop i\ j\ri\}$ (Def. \ref{eleSchles}), 
the difference of the Malgrange differential on the two solutions is  a {\bf closed differential on the deformation manifold} given by 
\be
\omega_{_M}(\pa;[\wt \Gamma]) - \omega_{_{M}} (\pa; [\Gamma]) = 
\pa \ln H\le\{\xi\ \eta\atop i\ j\ri\}
\ee
where 
\be
H \le\{\xi\ \eta\atop i\ j\ri\}= \le\{
\begin{array}{ll}
(\Gamma^{-1}(\xi)\Gamma'(\xi))_{ij} & \hbox { for  } \le\{\xi\ \xi\atop i\ j\ri\}\ (i\neq j)\\[14pt]
(\Gamma(\eta))_{ij} & \hbox { for  } \le\{\infty\ \eta\atop i \ j \ri\},\ \eta\neq \infty\\[14pt]
(\Gamma(\xi)^{-1})_{ij} & \hbox { for  } \le\{\xi\ \infty\atop i\ j \ri\} ,\ \xi \neq\infty\\[14pt]
\ds \frac{(\Gamma^{-1}(\xi)\Gamma(\eta))_{ij}}{\xi-\eta} & \hbox { for } \le\{\xi\ \eta \atop i\ j\ri\} \ , \ \xi\neq \infty\neq \eta
\end{array}
\ri.
\label{taudets}
\ee
where the notation $\Gamma'(\infty)$ --as previously-- denotes the derivative in the local parameter
\be
\Gamma(z)  =:\1 + \frac 1 z \Gamma'(\infty)+ \dots
\ee
\et
For the reader's convenience we verify Thm. \ref{tauratio} for the case $\le\{ \xi\ \eta\atop i\ j\ri\}$ with distinct $\xi\neq \infty\neq \eta$ and both $\xi,\eta\not \in \Sigma\gamma$.
Let $\pa$ be a variation of the jump-matrices $M$ (i.e. not moving $\xi,\eta$). We have from Prop. \ref{elemSchles}  
\be
R^{-1} R' = \frac {\Gamma(\eta)E_{ji} \Gamma^{-1}(\xi)}{(\Gamma^{-1}(\xi)\Gamma(\eta))_{ij}} \le(\frac 1 {z-\eta} - \frac 1{z-\xi}\ri)
\label{Rprime}
\ee
We then have to compute the (possibly formal) residue (\ref{deltaomega}) 
\bea
\res{z=\xi,\eta} \tr\le\{ \frac {\Gamma(\eta)E_{ji} \Gamma^{-1}(\xi)}{(\Gamma^{-1}(\xi)\Gamma(\eta))_{ij}} \le(\frac 1 {z-\eta} - \frac 1{z-\xi}\ri) \pa \Gamma(z)\Gamma^{-1}(z)\ri\} = \nonumber \\[6pt]
=\frac {(\Gamma^{-1}(\xi)\pa\Gamma(\eta))_{ij}}{(\Gamma^{-1}(\xi)\Gamma(\eta))_{ij}} 
- \frac { (\Gamma^{-1}(\xi)\pa\Gamma(\xi)\Gamma^{-1}(\xi) \Gamma(\eta))_{ij}}{(\Gamma^{-1}(\xi)\Gamma(\eta))_{ij}} =\nonumber \\[6pt]
=\frac {(\Gamma^{-1}(\xi)\pa\Gamma(\eta))_{ij}}{(\Gamma^{-1}(\xi)\Gamma(\eta))_{ij}} 
+  \frac {(\pa(\Gamma^{-1}(\xi) ) \Gamma(\eta))_{ij}}{(\Gamma^{-1}(\xi)\Gamma(\eta))_{ij}}=\nonumber \\[6pt]
=  \pa \ln \le( \frac{(\Gamma^{-1}(\xi)\Gamma(\eta))_{ij}}{\xi-\eta}
 \ri) 
\eea
To verify the formula for $\pa_\xi, \pa_\eta$ as well we must add a small circle around them and  a new jump $M(z) = (z-\xi)^{E_i}$, $M(z) = (z-\eta)^{-E_j}$ respectively. Since $\xi, \eta$ did not exist as deformation parameters in the RHP for $\Gamma$, from the definition of $\omega_{_M}$ we need to compute (we do it only for $\pa_\xi$, leaving the verification for $\pa_\eta$ to the reader)
\bea
\omega_{_M}(\pa_\xi;[\wt \Gamma]) = -\oint_{|z=\xi|=\epsilon} \tr \le( \wt \Gamma_-^{-1} \wt \Gamma_-' \frac {E_i}{x-\xi} \ri)\dd.
\eea
Note that $\wt \Gamma_- = R\Gamma$ (and $\wt \Gamma_+ = R \Gamma\, (z-\xi)^{-E_j}$) so that 
\bea
\omega_{_M}(\pa_\xi;[\wt \Gamma]) =-\res{z=\xi} \tr \le(\frac { R^{-1}(z)R'(z) \Gamma (z)E_i \Gamma^{-1}(z)}{z-\xi} + \frac {\Gamma^{-1}(z) \Gamma'(z) E_i}{z-\xi}\ri) =\cr
=  -\frac 1{\xi-\eta} - \frac {\big(\Gamma^{-1}(\xi)\Gamma'(\xi)\Gamma^{-1}(\xi)\Gamma(\eta)\big)_{ij}}{
(\Gamma^{-1}(\xi) \Gamma(\eta))_{ij}} = \pa_\xi \ln \le( \frac{\big (\Gamma^{-1}(\xi)\Gamma(\eta)\big)_{ij}}{\xi-\eta}
\ri)
\eea
This proves completely the case considered. 
It appears quite obviously that 
\bp
An elementary Schlesinger transformation  exists if and only if $H\le\{\xi\ \eta \atop i\ j\ri\} \neq 0$. 
\ep
\br
The zeroes of the matrix entries of the solution of the RHP acquire therefore the meaning of intersection of the Malgrange divisor with space of the parameter added to the problem, namely, the position of the Schlesinger transform $\le\{\infty\ \eta\atop i\ j\ri\}$.
\er
We are not going to dwell at length on the algebra of iterated elementary Schlesinger transformations and on the general transformation since the formul\ae\ are contained in \cite{JMU2}; we only point out that in loc. cit. the transformations were applied to solution of either isomonodromic or isospectral deformation problems and not a general Riemann--Hilbert problem. Thus, we are mainly shifting the perspective (and compactifying some notation) of \cite{JMU2}.
\br
The fact that $\omega$ evaluated on two solution of RHPs with the same jumps is a closed differential is immediate from the fact that the curvature of $\omega$ does not depend on the growth behavior of the solution $\Gamma$ but only on the jump matrices. When we  specialize the setting to the case relevant to ODEs, Thm. \ref{tauratio} will hold for differentiations with respect the monodromy data as well.
\er
\subsubsection{Generalized Sato formula}
\label{remSato}
Let $\eta\not\in \Sigma \gamma$:  denote by $\Gamma_{\eta;ij}(z)$ the  Schlesinger transform  $\le\{\infty \ \eta\atop i\ j\ri \}$ of the solution of the RHP \ref{probRH}. Then the second formula (\ref{taudets}) reads
\be
\Gamma_{ij}(\eta) \propto {\exp}{\int^{\vec s}\omega_{_M}(\bullet;[\Gamma_{\eta;ij}]) -\omega_{_M}(\bullet;[\Gamma])}  
\ee
where the one form under integration is closed by the above remark.
This is nothing but Sato formula for the Baker Akhiezer vector; of course, at this level of generality we do not have a ``tau'' function because --in general-- $\omega_{_M}$ will not be a closed differential of the deformation parameters of the problem. {\bf If} the problem admits a $\tau$--function, that is, {\bf if} the differential $\omega_{_M}$ (or $\Omega$) is closed on the submanifold of Riemann--Hilbert problems under consideration then we have a honest version of Sato formula
\be
\Gamma_{ij}(\eta) = \frac {\tau\le\{\infty \ \eta\atop i \ j\ri\}}{\tau}
\ee
where $\tau\le\{\infty \ \eta\atop i \ j\ri\}$ stands for the $\tau$--function of the problem with the ``insertion'' of the Schlesinger transform.

To make the remark a bit more concrete, let us pick a point $z=a\not\in\Sigma\gamma$ and a small disk $\mathbb D$. For simplicity in writing the formul\ae\ we will simply set $a=0$. 
Let  $T(z)$ be a diagonal matrix of the defined on $\pa\mathbb D$ that admits analytic continuation on $\mathbb P^1\setminus \mathbb D$
\be
T(z) = \sum_{k=1}^\infty \frac {T_k}{z^k}\ ,\  \ T_k= {\rm diag}( t_{k;1},\dots, t_{k,r})\ ,\label{diagonals}
\ee
where the Laurent series is supposed to be actually convergent on $\mathbb P^1\setminus \mathbb D$.
Let $\Gamma(z;[T])$ denote the solution of the RHP
\bea
\Gamma_+ = \Gamma_- M\ ,\ \ z\in \Sigma\gamma\\
\Gamma_+ = \Gamma_- {\rm e}^{T} \ ,\ \ z\in \pa\mathbb D\\
\Gamma(\infty)=\1
\eea
If the $L^\infty\cap L^2$ norm of ${\rm e}^{T}-\1$ is sufficiently small, then the solvability of the problem is guaranteed by standard perturbation theorems. Therefore $\Gamma(z;[T])$ is defined at least in a ball around $T=0$.  
Since the jump matrix ${\rm e}^{T}$ is diagonal (and analytic in the complement of the disk), it is immediately seen  from Prop. \ref{propcurv} that  $\omega_{_M} = \Omega$\footnote{This follows by observing that $\vartheta$ will be identically zero due to the analyticity of $T$ outside of the disk.} and that they are closed as differentials on the (infinite dimensional) manifold of $T$'s. Thus there is a locally defined function such that 
\be
\delta \ln \tau(T):= \oint_{\pa \D} \tr \le(\Gamma^{-1}_- \Gamma_-' \delta T \ri)\dd
\ee
Denote now by $\tau\le\{\xi\ \eta\atop i \ j \ri\}(\vec T)$\footnote{The notation $\vec T$ stands for the (infinite) vector of the (matrix) coefficients of $T(z)$, $\vec T = (T_1,T_2,\dots)$.} the tau--function resulting after the elementary Schlesinger transformation $ \le\{\xi\ \eta\atop i \ j \ri\} $. 
Then the content of Thm. \ref{tauratio} can be rephrased as 
\be
\Gamma_{ij}(\zeta) =\frac { \tau\le\{\infty\ \zeta\atop i \ j \ri\}(T) } {\tau(T)}\ ,\ \ {\Gamma^{-1}}_{ij}(\zeta) =\frac { \tau\le\{ \zeta\ \infty\atop i \ j \ri\}(T) } {\tau(T)}\
\ee
In fact, more is true: if $\zeta$ falls within the disk $\mathbb D$ then the RHP for the Schlesinger transforms $ \le\{\infty\ \zeta\atop i \ j \ri\}, \le\{\zeta\ \infty\ \atop i \ j \ri\} $ can be formulated as ({\bf exercise})
\be
\begin{array}{c|c}
\le\{\infty\ \zeta\atop i \ j \ri\} &  \le\{\zeta\ \infty\ \atop i \ j \ri\} \\
\hline
\Gamma_+= \Gamma_-{\rm e}^T \le(1-\frac \zeta z\ri)^{-E_j} z^{E_i-E_j}\ ,\ \ z\in \pa \D &\Gamma_+= \Gamma_-{\rm e}^T (1-\frac \zeta z)^{E_i} z^{E_i-E_j}\ ,\ \ z\in \pa \D \\
\Gamma_+ = \Gamma_- z^{-E_i} M z^{E_i} \ z\in \Sigma\gamma&\Gamma_+ = \Gamma_- z^{E_j} M z^{-E_j} \ z\in \Sigma\gamma \\
\Gamma(\infty)=\1&\Gamma(\infty)=\1
\end{array}
\ee
The jumps on $\pa\D$ can be written
\be
{\rm e}^{T(z)} \le(1-\frac \zeta z\ri) ^{-E_j} = \exp \sum_{k=1}^\infty \frac {T_k + E_{j} \zeta^k/k} {z^k}\ ;\ \ 
{\rm e}^{T(z)} \le(1-\frac \zeta z\ri) ^{E_i} = \exp \sum_{k=1}^\infty \frac {T_k - E_{i} \zeta^k/k} {z^k}
\ee
This leads to the following identities
\be
\Gamma_{ij}(\zeta;[T]) = \Gamma_{ij}\le(0;\le[T- E_j [\zeta]\ri]\ri)\ ,\ \ {\Gamma^{-1}}_{ij}(\zeta;[T]) = \Gamma_{ij}\le(0;\le[T+ E_i [\zeta]\ri]\ri)
\ee
and hence 
\be
\Gamma_{ij} (\zeta;[T]) = \frac { \tau\le\{ \infty\ \ 0\atop i \ j\ri\}(T-E_j[\zeta])}{\tau(T)}
\ ;\ \ \ 
\Gamma_{ij} (\zeta;[T]) = \frac { \tau\le\{ 0\ \ \infty\atop i \ j\ri\}(T+E_i[\zeta])}{\tau(T)}\label{273}
\ee
Here we have used the standard notation $[\zeta] = (\zeta, \frac {\zeta^2}2,\dots, \frac {\zeta^k}{k},\dots)$.
\subsubsection{Hirota bilinear relations}
\label{secHirota}
We will not go into much depth here, since all is well--known but it may give an {\em analytic}  perspective on the relations, which are usually taken only formally.

As in the previous Section \ref{remSato} let $\mathcal C$ be a counterclockwise circle around a point $z=a\not\in \Sigma \gamma$  ($a=0$ for simplicity) and let $T(z):\mathcal C\to  gl_r(\C)$ be as (\ref{diagonals}). 
We have 
\be
\oint_{\mathcal C} \Gamma_+(\zeta;[T]) {\rm e}^{-T(\zeta)+ \wt T(\zeta)} \Gamma^{-1}_+(\zeta;[\wt T])\frac {\d \zeta}{2i\pi \zeta^2} = \oint_{\mathcal C} \Gamma_-(\zeta;[T])  \Gamma^{-1}_-(\zeta;[\wt T])\frac {\d \zeta}{2i\pi \zeta^2}
\ee
Now, the matrix $ \Gamma(x;[T])  \Gamma^{-1}(x;[\wt T])$ has {\bf no jumps outside $\mathcal C$} since the other jumps of the problem have been left unmodified; thus it is analytic on the complement of the disk $\mathbb D$ and goes to the identity at $\infty$. Thus we have the identity
\be
\oint_{\mathcal C} \Gamma_+(\zeta;[T]) {\rm e}^{-T(\zeta)+ \wt T(\zeta)} \Gamma^{-1}_+(\zeta;[\wt T])\frac {\d \zeta}{2i\pi \zeta^2} =  0 \label{hirota}
\ee
which is valid {\em identically} in $T, \wt T$ in a neighborhood of  $T\equiv 0$.

In view of the interpretation of $\Gamma$ as the Baker--Akhiezer vector for the $\tau$--function (Sato-formula above), the reader may regard (\ref{hirota}) as the generating function of an infinity of bilinear identities between  matrix--valued $\tau$ functions;
\be
\sum_{\ell=1}^r\oint  \tau\le\{ \infty\ \ 0\atop i \ \ell\ri\}(T-E_\ell [\zeta]){\rm e}^{\wt T_\ell(\zeta)- T_\ell(\zeta)}  \tau\le\{ 0\ \ \infty\atop \ell \ j\ri\}(\wt T+E_\ell[\zeta]) \frac {\d \zeta}{\zeta^2} \equiv 0 \label{Hirotaaa}
\ee
The identity (\ref{Hirotaaa}) should be used as a generating function of an infinite hierarchy of PDEs for the matrix--valued tau-function $[\boldsymbol\tau(T)]_{ij} = \tau\le\{\infty \ 0\atop i\ j\ri\}(\vec T)$ when expanding it in Taylor series with respect to $\wt T$ around the diagonal $\wt {\vec T} =\vec  T$.
This generates a sort of ``addition theorem'' for tau-functions (see Remark 2 in \cite{JMU3}).

The variational formula (\ref{pakern}) in Lemma \ref{varGamma} takes on an added significance in view of the identity
\be
 \frac 1{\tau} \tau\le\{ \xi\ \eta\atop i\ j \ri\}  =\le[\frac {\Gamma^{-1}(\xi)\Gamma(\eta) }{\xi-\eta}\ri]_{ij} 
\ee
To explain this in a simple situation we now consider {\bf two disks} $\mathbb D_0$ and $\mathbb D_1$ centered at two points --say-- $a=0,1$; on the boundaries of these  disks we introduce diagonal jumps exactly as in Sec. \ref{remSato}
\be
T^{(0)} (z) = \sum_{k=1}^{\infty} \frac {T^{(0)}_k}{z^k}\ ,\ \ \T^{(1)} (z) = \sum_{k=1}^{\infty} \frac {T^{(1)}_k}{(z-1)^k}
\ee  
We denote by $\tau(\vec T^{(0)}, \vec T^{(1)})$ the tau--function as a function of the two (infinite) sets of times and $\tau\le\{0\ 1\atop i\ j\ri\}(\vec T^{(0)} , \vec T^{(1)})$ the Schlesinger--transformed one.
Let $\zeta\in \mathbb D_0$ and $\eta\in \mathbb D_1$; retracing the steps that lead to (\ref{273}) we find 
\be
\frac 1{\tau(\vec T^{(0)}, \vec T^{(1)})}\tau\le\{0\ 1\atop i\ j\ri\}(\vec T^{(0)} + E_i[\xi], \vec T^{(1)} - E_j[\eta]) =  \le[\frac {\Gamma^{-1}(\xi)\Gamma(\eta) }{\xi-\eta}\ri]_{ij}\label{JMU311}
\ee
This formula is the content (in different notation) of (3.11) in Thm. 3.2 of \cite{JMU3}.

The variational equation (\ref{pakern}) in our Lemma \ref{varGamma} applied to derivatives with respect to some directions in $T^{(0)}$, $T^{(1)}$ then become the generating functions for the Hirota bilinear relations that appear in Thm. 3.4 of \cite{JMU3}.
They all boil down to the following identity, to be understood as generating functions of PDEs when evaluating its Taylor expansion on the diagonal $T^{(j)} = \wt T^{(j)}$, $j=0,1$
\bea
0 =\le( \oint_{\pa \mathbb D_0} + \oint_{\pa \mathbb D_1} \ri)\frac {\Gamma^{-1}(\xi; T) \Gamma_-(x;T)}{\xi-x} \frac {\Gamma_-^{-1} (x;\wt T) \Gamma(\eta;\wt T)}{x-\eta} \dd = \nonumber \\[12pt]
\oint_{\pa \mathbb D_0}\frac {\Gamma^{-1}(\xi;T) \Gamma_+(x;T)}{\xi-x} {\rm e}^{\wt T^{(0)}-T^{(0)}} \frac {\Gamma_+^{-1} (x;\wt T) \Gamma(\eta;\wt T)}{x-\eta} \dd   + \oint_{\pa \mathbb D_1}\frac {\Gamma^{-1}(\xi;T) \Gamma_+(x;T)}{\xi-x} {\rm e}^{\wt T^{(1)}-T^{(1)}} \frac {\Gamma_+^{-1} (x;\wt T) \Gamma(\eta;\wt T)}{x-\eta} \dd
\eea
The interested reader should compare this with  Theorem 3.4 in \cite{JMU3}.
Since it is not the primary focus of this paper (and it is certainly not a new result), we will not pursue the issue here, also because it has been dealt with at length in \cite{JMU3}, even though in the context of isomonodromic and isospectral deformations only.
\subsection{Right gauge equivalence}
\label{secGauge}
We will say that the two problems are {\bf (right) gauge equivalent} if there exists an analytic function 
\be
G:\C\setminus \Sigma\gamma \to GL_r(\C)
\ee
admitting boundary values (also for its derivative) at $\Sigma\gamma$ and such that the jump matrices stand in the relation
\be
\wt M (z) = G_-^{-1}(z) M(z) G_+(z)\ ,\ \ z\in \Sigma \gamma.
\ee
It is immediate then that the two solutions are related by $
\wt \Gamma(z) = \Gamma(z) G(z)$.
It is then seen that the difference of $\omega_{_M}$ (or $\Omega$) along the two solution $\Gamma, \wt \Gamma$ differ only in terms that do not involve $\Gamma$ or $\wt \Gamma$ and depend only and explicitly on $M, G$. Thus this equivalence will not modify the singularity locus of $\omega_{_M}$ and --if both deformation families admit a tau function-- both tau functions will differ only by multiplication by a smooth nonzero factor. Indeed a direct computation yields 
\bea
\omega_{_{M}}(\pa;[\wt \Gamma]) - \omega_{_{M}} (\pa;[\Gamma]) = \int_{\Sigma\gamma} \tr 
\le(\Gamma_-^{-1} \Gamma_-' G_-^{-1} \pa  G_- -\Gamma_+^{-1} \Gamma_+' G_+^{-1} \pa  G_+ \ri) \dd +\label{zero} \\
+ \int_{\Sigma\gamma} \tr \le( M^{-1} M' G_+^{-1} \pa G_+ - G_-^{-1} G_-' G_-^{-1}\pa G_-  - G_-^{-1} G_-' \pa M M^{-1} + G_-^{-1} G_-' M G_+^{-1} \pa G_+ M^{-1} \ri)\dd
\eea
The term in (\ref{zero}) vanishes by Cauchy theorem, since it amounts to the (boundary value of) the integral of $\tr (\Gamma^{-1} \Gamma' G^{-1} \pa G)$ on a collection of contours surrounding $\Sigma \gamma$ and contractible. Thus the difference will not involve the solution of the RHP and hence be a smooth differential in the parameters (i.e. gauge equivalence cannot modify the Malgrange divisor). Such equivalence does play a role in some cases (see \cite{Bertola:MomentTau} for examples where this happens, although not phrased in these terms).
\section{Rational differential equations in terms of Riemann--Hilbert data}
\label{RHPODE}
We now describe the class of ODE's with rational coefficients
\be
\Psi'(z) = A(z) \Psi(z)
\ee
in an unconventional way: we will start from the formulation of a RHP and then indicate how this problem (when solvable) is equivalent to an ODE.
The class of  matrices $A(z)$ that will be eventually described has poles at points $a_1,\dots, a_k$ with orders $n_1+1,\dots, n_k+1$. If $n_j=0$ then the pole is simple. 

One may take the point of view that we are providing a different (transcendental) coordinate system on the finite--dimensional vector space of rational matrices with fixed polar divisor.

The forward problem, namely, the construction of the RHP from the matrix $A(z)$ is more standard and we only sketch the main points, since it does not really play a direct r\^ole here. This procedure is often called the {\bf (extended) monodromy map}. The standard reference for many assertions below is Wasow's book \cite{Wasow} but also the paper \cite{JMU1} provides a concise recall.

\subsection{Monodromy map}
\label{Birk2}
Given a rational matrix 
\be
A(z) = \sum_{j=1}^K \sum_{\ell=1}^{n_j+1} \frac {A_{j,\ell-1}}{(z-a_j)^\ell} + \sum_{\ell=0}^{n_0-1} A_{0,\ell+1} z^\ell 
\ee
we consider the ODE $\Psi'(z) = A(z)\Psi(z)$. Without loss of generality we will assume $\tr A(z) \equiv 0$ so that any solution has constant determinant that we can assume to be unity $\det \Psi\equiv 1$.
We make the usual assumption that 
\begin{assumption}[Genericity]
\label{genericity} The leading-coefficient matrices $A_{j,n_j}$ have distinct simple eigenvalues. The diagonal matrix of eigenvalues (with an arbitrarily chosen order) will be denoted by $T_{j,n_j}$. In addition, if $n_j=0$ (simple pole) then $T_{j,0}$ has eigenvalues which are also distinct modulo $\Z$ (i.e. no pair of eigenvalues differ by an integer).
\end{assumption}

To simplify some issues in the general description  we will assume that in fact there is no pole of the connection $\pa_z - A(z)$ at $z=\infty$; this can always be achieved without loss of generality by a M\"obius transformation that maps $\infty$ to a finite point (without mapping any of the other poles to infinity!). This allows to choose as basepoint $z_0$ the point at infinity.

\paragraph{Monodromy representation.} We choose a basepoint $z_0$ or the homotopy group and consider the initial value problem $\Psi(\infty )=\1$. By analytic continuation  of the solution around a loop that  ``goes around $a_j$'' (i.e. has index one relative to $a_j$ and zero relative to all other poles) we obtain $\Psi(z)\mapsto \Psi(z) M_j^{-1}$, $\det M_j=1$. These loops generate the fundamental group $\pi_1(\C \setminus \{a_1,\dots\}, z_0)$ and provide a representation of this fundamental group 
\be
\pi(\mathbb P^1 \setminus \{a_1,\dots\}, z_0) \mapsto SL_r(\C).
\ee
Note that the basepoint for the normalization ($\infty$ in our case) and the basepoint for the homotopy group may not be the same. We will denoted by $\mathcal D$ the simply connected domain of $\mathbb P^1$ obtained by dissecting $\mathbb P^1$ along nonintersecting smooth arcs joining $z_0$ with each of the poles.
\paragraph{Stokes' phenomenon in brief.} Consider a higher order pole $a_j$, with $n_j\geq 1$ and denote as follows the distinct eigenvalues of the leading coefficient matrix $A_{j,n_j}$
\bea
T_{j,n_j} = {\rm diag} (\Lambda_1,\dots, \Lambda_r)\ ,\ \ \Lambda_i\neq \Lambda_j, \ \ i \neq j\\
\zeta:=(z-a_j)\ ,\ \ n:=n_j\ ,
\eea
(or $\zeta = \frac 1 z$ for the pole at infinity).
One can find at each pole $a_j$ $2n_j$ directions (the {\bf anti-Stokes} directions) angularly separated by $\frac {\pi}{n_j}$ such that along each o f them there exists a permutation $\sigma$ (uniquely defined) yielding the definite ordering below 
\be
\Re \le(\frac {\Lambda_{\sigma(1)}}{\zeta^n}\ri) > \Re \le(\frac {\Lambda_{\sigma(2)}}{\zeta^n}\ri) > \dots \Re \le(\frac {\Lambda_{\sigma(r)}}{\zeta^n}\ri) \label{order}
\ee
Note that if on a direction we have the above ordering, on the next (counter)clockwise we have the exact reversed (with the same permutation). 

The theorem which can be found in several place (e.g. \cite{Wasow}) is the following; given our IVP $\Psi(z)$ on the simply connected domain $\mathcal D$, there are $2n$ matrices $S_\nu\ ,\ \nu=1\dots 2n$ (the {\bf Stokes' matrices}) which are  of the form 
\be
S_\nu = P_\sigma (\1 + N_\nu) P_\sigma^{-1}
\ee
where $N_\nu$ is upper-triangular on the directions where the ordering is as in (\ref{order}) and lower-triangular on the directions where the ordering is reversed, and $P_\sigma$ is the permutation matrix corresponding to the permutation $\sigma$ appearing in (\ref{order}). 
There is an invertible matrix $C=C_j\in SL_r(\C)$ (the {\bf connection matrix}) and {\bf diagonal} matrix 
\be
H(\zeta):= {\rm e}^{T(\zeta)} \zeta^{L}\ ,\ \ 
T(\zeta) = \sum_{\ell=1}^{n} T_\ell \zeta^{-\ell}\ ,\ \ \zeta = (z-a) \hbox { (or $\zeta = \frac 1 z$ for the pole at $\infty$})
\ee
in which $T_n= T_{j,n_j}$ is the matrix of eigenvalues of the leading coefficient used in the definition of the anti-Stokes directions. This matrix will be called the {\bf toral element} (for lack of a better naming, since it belongs to the complex toral subalgebra of $SL_r(\C)$) and the matrix $L$ is called the {\bf exponents of formal monodromy}.

These matrices (Stokes' connections and toral elements) are uniquely determined by the following set of conditions
\begin{enumerate}
\item In one (arbitrarily chosen and then fixed) sector separated by the two consecutive anti-Stokes directions we have the asymptotic expansion
\bea
\Psi(z) &\& \sim \wh Y(z)  {\rm e}^{T(\zeta)} \zeta^{L} C \\
&& \wh Y(z) = G\le(\1 + \sum_{j=0}^{\infty} Y_j \zeta^j\ri)  , \det G=1
\eea
\item In the next sector counterclockwise we have 
\be
\Psi(z)  \sim \wh Y(z)  {\rm e}^{T(\zeta)} \zeta^{L} S_1 C 
\ee
where we have labeled by $1$ the anti-Stokes direction separating the two sectors, and in general
\be
\Psi(z) \sim \wh Y(z)  {\rm e}^{T(\zeta)} \zeta^{L}S_\nu\cdots S_2\cdot S_1 C
\ee 
where we are in the sector between two anti-Stokes and we have crossed $\nu$ such lines.

\item The monodromy around the pole under consideration, the connection matrix and the Stokes matrices satisfy the condition 
\be
M = C^{-1}  S_{2n} \cdots S_1 C
\ee
\end{enumerate}
Note that at a simple pole we simply have $n=0$ and hence there are no anti-Stokes' lines but only the exponents of formal monodromy $L$ and the connection matrix $C$.

\br[\bf Notational issue]
It should be understood that all the above matrices ($C, S, T,L$) are different for each of the poles and hence we should understand an index distinguishing them and related to the pole under consideration.
\er
\br[\bf Important]
The triangularity condition for the Stokes' matrices $S_\nu$ can be expressed intrinsically by saying that 
\be
M_\nu(z):= {\rm e}^{T(z)} S_\nu {\rm e}^{-T(z)} = \1 + \mathcal O( (z-a)^\infty)
\ee
as $z$ approaches $a$ along the corresponding anti-Stokes direction.
\er

Since these are standard facts about ODEs we will not dwell on other subtleties.

\subsubsection{Forward Birkhoff map.} Given the rational matrix $A(z)$ of our form, a  chosen basepoint and dissection $\mathcal D$, anti-Stokes lines and base-sector from where to start the counting we have associated the collection of all the data 
\be
\mathcal M := \{a_j, T_j(\zeta_j),L_j,  C_j, \{S_{\nu,j} \nu =1\dots 2n_j\},\}_{j=1,\dots, k}
\ee
The statement is that the map is (locally) injective but not surjective in general; there are some choices of data in $\mathcal M$ for which there is no corresponding ODE of the specified form. As the reader may suspect (or know), these data constitute the Malgrange divisor.

There are two logically (and historically) distinct types of data in the above: the {\bf isomonodromic times}
\be
\mathcal T:=  \{a_j, T_j(\zeta_j)\}_{j=1\dots k}
\ee
(the positions $a_j $ and the coefficients of the matrices $T_j(\zeta)$). On the other hand there are the genuine generalized monodromy data
\be
\mathcal S:= \{L_j,  C_j, \{S_{\nu,j} \nu =1\dots 2n_j\},\}_{j=1\dots k}
\ee
and $\mathcal M$ is locally the product of the two. Of course this separation can be only done locally since  the anti-Stokes directions depend on the leading coefficients of $T_j(\zeta)$ and the dissection $\mathcal D$ depends on the position of the poles,  and hence some care is in order --if global question are at issue-- in describing the patching of these local descriptions (see for example \cite{Boalch1}).
\section{Inverse Birkhoff map: Riemann--Hilbert problem}
\label{SecBirk}
Given the Birkhoff data described above, the question arises as to whether one can invert the map: starting from a concrete dissection, anti-Stokes lines etc, together with all the matrices appearing in $\mathcal M$,  can one {\em reconstruct} $A(z)$?. To this end it is necessary to specify a Riemann--Hilbert problem in the same spirit as Section \ref{RHPsec}.

\subsection{The set of contours} 
The set of contours $\Sigma \gamma$ consists of (see  Fig. \ref{fig1})
\begin{enumerate}
\item for each pole we draw a circle not containing any other pole: we will call this the {\bf connection circle};
\item for each pole $a_j$  a smaller circle is chosen, called the {\bf formal monodromy circle}. On this circle a point $\beta_j$ is chosen.
\item each $\beta_j$ on the formal monodromy circle connected with a set of mutually nonintersecting paths ({\bf stems})  to the basepoint;
\item at the higher poles we choose a third smaller circle called the {\bf toral circle};
\item a point (arbitrarily chosen) on the {\em toral circle} is connected finally to the pole by  $2n$ smooth curves that approach the singularity  along the directions mentioned in Section \ref{Birk2}; 
\end {enumerate}
The word description is awkward but Fig. \ref{fig1} should clarify all the elements.
\subsection{The jump-matrices}
\begin{figure}
\resizebox{0.91\textwidth}{!}{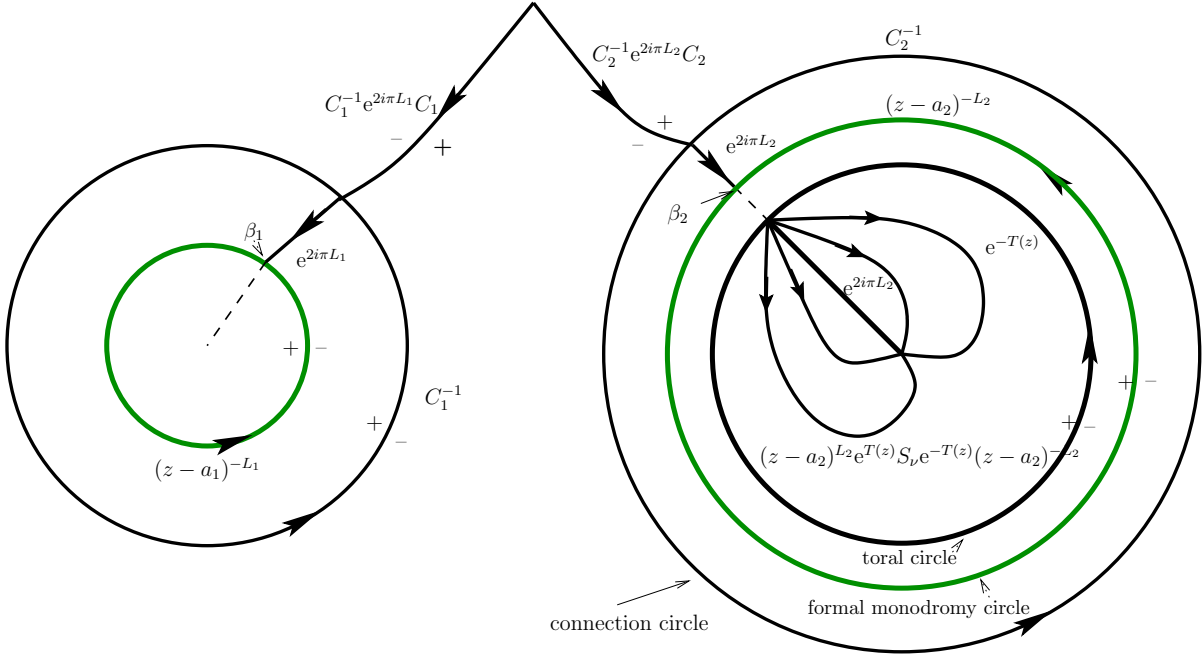}
\caption{The typical set of contours. The dashed lines within the formal monodromy circles are not jumps, but the domain where the determination of the powers $(z-a)^{-L}$ is continuously defined so as to give a precise meaning to the expressions for the various jump matrices.}
\label{fig1}
\end{figure}

Rather than describing the matrices in words, we refer to the picture (Fig. \ref{fig1}) where we depict a situation with only two points $a_1,a_2$ with $n_1=0$ (simple pole). The general picture is quite simply a repetition of several copies of the basic elements already manifest here. Since there is no monodromy around the basepoint $z_0=\infty$ we have to impose the constraint 
\be
C_n^{-1}{\rm e}^{2i\pi L_n} C_n \cdots C_1^{-1}{\rm e}^{2i\pi L_1} C_1 = \1
\ee
It will be understood that the matrices $(z-a)^L$ are defined on the formal monodromy circles as continuous functions taken away the point of insertion of the points $\beta_j$ (Fig. \ref{fig1}).

\subsection{Riemann--Hilbert problem}
\begin{problem}
Find a piecewise analytic matrix--valued function $\Gamma(z)$ on the complement of the indicated contours as per Fig. \ref{fig1} so that
\begin{itemize}
\item on each arc it solves $\Gamma_+(z) = \Gamma_-(z) M(z)$ with the matrix $M(z)$ as indicated in Fig. \ref{fig1};
\item it is bounded on $\mathbb P^1$;
\item satisfies the normalization $\Gamma(\infty)= \1$.
\end{itemize}
\end{problem}
Some properties follow immediately:
\begin{itemize}
\item Any solution satisfies $\det \Gamma(z)\equiv 1$. Indeed $\det M(z)\equiv 1$ implies that $\det (\Gamma(z))$ has no jumps across the contours. Since $\Gamma(z)$ is bounded so must be $\det \Gamma$ and hence it is an entire function, bounded everywhere, hence a constant. Since $\Gamma(z_0)=\1 $ then $\det \Gamma(z) = \det \Gamma(z_0)=1$.
\item If a solution exists, it is unique: indeed any two solutions $\Gamma, \wt \Gamma$ are analytically invertible by the above remark. Hence $\wt \Gamma \Gamma^{-1}$ is promptly seen to have no jumps, be uniformly bounded, hence a constant. The normalization forces $\Gamma\equiv \wt\Gamma$.
\end{itemize}

\begin{wrapfigure}{r}{0.4\textwidth}
\resizebox{0.4\textwidth}{!}{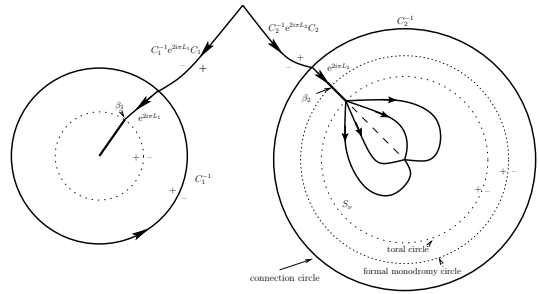}
\caption{The constant-jump Riemann--Hilbert problem for $\Psi$.}
\label{RHPPsi}
\end{wrapfigure}

The connection with  the  ODE is as follows; define the piecewise analytic matrix 
$\Psi(z)$ related to the solution $\Gamma$ as follows:
\begin{itemize}
\item $\Psi(z)=  \Gamma(z)$  outside the formal monodromy circles;
\item  $\Psi(z)  = \Gamma(z) (z-a_j)^{L_j}$ in the annulus between toral and formal-monodromy circles;
\item $\Psi(z) =  \Gamma(z) {\rm e}^{T_j(z)} (z-a_j)^{L_j}$  inside the formal-monodromy circles. 
\end{itemize}
The cut of the logarithm is taken where the dotted line is traced in Fig. \ref{fig1}.
It   is promptly seen that $\Psi(z)$  solves a new RHP where the jump matrices are piecewise constant and unimodular. Thus 
\be
A(z):= \Psi'(z) \Psi^{-1}(z)
\ee
is a meromorphic (traceless) matrix function with isolated singularities at the $a_j$'s. A local inspection shows that it has poles of finite order there and hence it is rational.
\section{Tau function and (iso)monodromic deformations}
\label{tauiso}
The goal now is to show that the modified Malgrange form $\Omega$ is closed when evaluated along the manifold $\mathcal L\subset \mathcal G$ consisting of jump-matrices $M(z)$'s described in the previous section.
We emphasize that the parameters are 
\begin{itemize}
\item the toral data $T_j(z)$'s;
\item the connection matrices $C_j$;
\item the Stokes' matrices $S_{\nu,j}$;
\item the exponents of  formal monodromy $L_j$.
\end{itemize}
The reader acquainted with the literature about isomonodromic deformations should now realize that we allow many more directions of deformations.
The computation of the closure of $\Omega$ relies directly on Prop. \ref{propcurv}:  
\be
\delta \Omega(\pa,\wt \pa) = \pa \Omega(\wt \pa)  - \wt \pa \Omega( \pa)  = \frac 1 2 \int_{\Sigma \gamma} \tr \bigg(M' M^{-1} \le[\pa M M^{-1} ,  \wt \pa M M^{-1}\ri]\bigg)\dd\label{curvomm}
\ee
It is clear that no contribution to (\ref{curvomm})  can come from the arcs where $M(z)$ is independent of $z$. This leaves only the contributions coming from the toral circles, the formal monodromy circles  and the Stokes' lines.
Let us consider each type separately
\begin{itemize} 
\item {\bf Stokes' lines}; up to a constant conjugation by a permutation matrix the jumps are of the form $\1 + N(z)$ with a strictly triangular $N(z)$, and hence they do not contribute to (\ref{curvomm}) (the trace is identically zero);
\item {\bf Toral circles}; on the Toral circles the jumps are diagonal and hence the commutator in (\ref{curvomm}) is identically zero. 
\item {\bf Formal monodromy circles}; once more, since the jumps are diagonal, no contribution to (\ref{curvomm}) comes from them. Note that at a simple pole the matrix $L$ may actually be upper triangular: in this case the trace vanishes identically.
\end{itemize}

Therefore we have the 
\bt
\label{main}
The differential $\Omega$ restricted to the submanifold $\mathcal L$ of group--valued jump-matrices described above is closed and defines a local function via the formula 
\be
\tau(\vec T, \vec a, \vec L, \vec S, \vec C) = {\rm e}^{\int \Omega}
\ee
This function is defined up to nonzero multiplicative constant and it vanishes precisely when the Riemann--Hilbert problem is not solvable, namely, on the Malgrange $\Theta$ divisor.
\et
The fact that $\tau$ has zeroes (and not a branching singularity) does not follow from our construction, it follows from \cite{Palmer:Zeros} and \cite{Malgrange:IsoDef1}.

We would like to stress that, had we chosen the Malgrange form $\omega_{_M}$ directly, the closure would have failed. 
 Indeed the reader may check that 
 \bea
 \delta \omega_{_M}\le(\pa , \wt \pa\ri) =\frac 1 2 \sum_j\le( \pa a_j \wt \pa \tr (L_j^2) -  \wt \pa a_j \pa \tr (L_j^2)\ri)K_j\\
 K_j = \oint_{\beta_j} \frac{\ln (x-a_j)}{(x-a_j)^2} \dd = \frac {1}{\beta_j-a_j} \neq 0
 \eea
where the integrals are on the toral circles with basepoint  at the necks $\beta_j$.

Aside from this, the correction therm $\vartheta$ in (\ref{recurv}) does not contribute anything for most of the contours in $\Sigma \gamma$; in fact the only contribution comes --not surprisingly in view of the above computation-- from the formal monodromy circles 
\bea
\vartheta(\pa) = \frac 12 \int_{\Sigma\gamma} \tr (M'M^{-1} \pa M M^{-1}) \dd = \frac 12  \sum_{j} \tr (L \pa L) \oint_{\beta_j}\frac {\ln (x-a)}{x-a}\dd =\\
= \frac 12  \sum_{j} \tr (L \pa L) (2i\pi + 2\ln(\beta_j-a_j)) 
\eea

Note that this ``problem'' is invisible if we allow only isomonodromic deformations, in which case using $\omega_{_M}$ or $\Omega$ yields the same differential.

 Before we turn to a list of applications of the above result to explicit examples, we make the connection with the definition of \cite{JMU1}.
\subsection{Relationship between $\omega_{_M}$ and $\omega_{_{JMU}}$}
\label{MalgrangeJMU}
\begin{wrapfigure}{r}{0.4\textwidth}
 \resizebox{0.4\textwidth}{!}{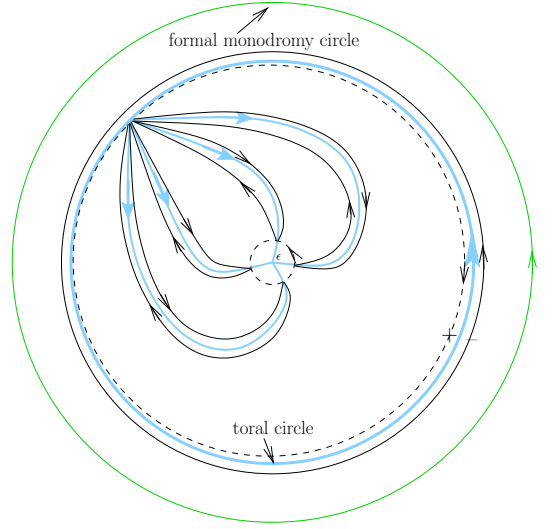}
 \caption{The set of contours for the computation of $\omega_{_{JMU}}$}
 \label{JMUfig}
 \end{wrapfigure}
In order to establish the relationship we must ``freeze'' all the monodromy part of the Birkhoff data, that is, the connection matrices, the Stokes matrices and the exponents of formal monodromy. As noted above, in this case $\Omega = \omega_{_M}$. 
Then the observation that $\omega_{_M}$ and $\omega_{_{JMU}}$ coincide can be dug out \cite{Palmer:Zeros} but we re-derive it here for the sake of self-containedness. If now $\pa$ is a derivative along the isomonodromic submanifold of the Birkhoff data, the differential reduces to an integral only on the Stokes' lines, the toral circles and the formal-monodromy circles. Since the expression is repeated for each pole, we consider only the localization at one pole $a=a_j$, i.e. we vary the toral data/position only of one pole. Let $\pa$ denote one such deformation involving only data at $a=a_j$. Then 
\bea
\Omega(\pa) = \omega_{_{M}}(\pa) =- \int \tr (\Gamma_-^{-1} \Gamma_-' \pa M M^{-1}) \dd
 \eea

 There are two types of integrals here: the integrals along the Stokes lines and the integral around the toral circle.
 Along each Stokes contour the integrand can be equivalently written as follows 
 \bea
 \tr \le(\Gamma_-^{-1} \Gamma_-' \le(\pa \wh T - \overbrace{{\rm e}^{\wh T}S {\rm e}^{-\wh T}}^{M}\pa\wh T
 \overbrace{{\rm e}^{\wh T} S^{-1} {\rm e}^{-\wh T}}^{M^{-1}}\ri) \ri)  =\cr=  \tr \le(\Gamma_-^{-1} \Gamma_-' \pa\wh  T - \Gamma_+^{-1} \Gamma_+' \pa \wh T\ri) + \cancelto{0 (\hbox{\tiny traceless})}{\tr(M^{-1} M' \pa \wh T)}  
\label{JJJ} \\
\wh T(z):= T(z) + L \ln(z-a)
\eea
where we have used the cyclicity of the trace on the second term followed by  
\be
M^{-1}\Gamma_-^{-1} \Gamma_-'  M  = \Gamma_+^{-1} \Gamma_+' - M^{-1} M' 
\ee
Recall also that expressions like $\pa \wh T$ do not contain a logarithmic term because $\pa L =0$ (only isomonodromic deformations are allowed).
As a consequence of (\ref{JJJ}) the resulting sum of integrals (refer to Fig. \ref{JMUfig}) can be evaluated by excising an $\epsilon$ circle\footnote{Note that $\Gamma$ has a jump of the form ${\rm e}^{2i\pi L}$ in the straight ray from $a$ to the toral circle (Fig. \ref{fig1}), but the expression $\tr (\Gamma^{-1} \Gamma' \pa \wh T)$ does {\bf not have a jump} there, since $L$ and $\wh T$ are both diagonal. This is the reason why we did not draw the corresponding solid line in Fig. \ref{JMUfig}.} around the pole $z=a$ and integrating $\tr (\Gamma^{-1} \Gamma' \pa T)$ on the solid black contours indicated in Fig. \ref{JMUfig}, followed by a limit $\epsilon \to 0$. For fixed $\epsilon$ the integral along the solid ``tentacles'' is equal (by Cauchy's theorem) to the  counterclockwise integral along the $\epsilon$ circle  and the clockwise integral along the bigger dashed circle indicated in Fig. \ref{JMUfig}. 

The two integrals  of  $\tr (\Gamma^{-1} \Gamma' \pa T)$ along the two circles surrounding the toral circle reduce --due to the jump of $\Gamma$ there to 
\be
-\oint_{toral} \tr (\Gamma_-^{-1} \Gamma_-' \pa T- \Gamma_+^{-1} \Gamma_+' \pa\wh  T)\d x =\cancelto{0}{-\oint_{toral} \tr ( \wh T'\pa\wh  T)} +\oint_{toral} \tr\le(\Gamma_-^{-1} \Gamma_-' \frac {L\pa a}{x-a}\ri)\dd
\ee
where the first integral is zero because it is residueless. The remaining can be deformed within the annulus between the toral and formal-monodromy circle (and the $\Gamma_-$ of the toral circle becomes the $\Gamma_+$ of the local monodromy circle) and combines with the outer integral therein to give 
\be
\oint_{for.mon.} \tr \le(\le (\Gamma_+^{-1} \Gamma_+' - \Gamma_-^{-1} \Gamma_-' \ri )\frac {L\pa a}{a-x}\ri)\dd = - \oint_{for.mon.}  
\frac{\tr(L^2) \pa a^2}{(z-a)^2}\dd =0
\ee
The remaining integrals can therefore be retracted to an integral around the $\epsilon$ circle:
\be
\oint_{|z-a|=\epsilon} \tr \le(\Gamma^{-1} \Gamma' \pa \wh T\ri)\dd\label{ecirc}\ .
\ee 
In each sector we have  the expansion valid for any $N$
\be
\Gamma(z) = G \le(\1 + \sum_{0}^{N} Y_k (z-a)^k\ri) + \mathcal O((z-a)^{N+1}) \sim   G \le(\1 + \sum_0^\infty Y_k (z-a)^k\ri) 
\ee
where the coefficient matrices $G, Y_k$ are the same irrespectively of the sector. If we replace $\Gamma(z)$ in (\ref{ecirc})  by a suitably high truncation of the formal series we commit an error that it is easily estimated to vanish as $\epsilon$ tends to zero.
On the other hand the new integral is independent of $\epsilon$ and reduces to the formal residue
\be
\Omega(\pa;[\Gamma]) = \omega_{_M}(\pa;[\Gamma]) = \res{z=a} \tr \le(\wh Y^{-1}(z)  \wh Y'(z)\pa \wh T(z) \ri)\ ,\ \ \ \wh Y := G \le(\1 + \sum_0^\infty Y_k (z-a)^k\ri)
\ee 
to be understood simply as  the coefficient of $(z-a)^{-1}$ of the above formal series with a {\em finite} Laurent tail.
The reader acquainted with the definition of $\omega_{JMU}$ will recognize that the expression is precisely the same defining $\omega_{_{JMU}}$ in \cite{JMU1} (of course one should repeat the residue computation at each pole and sum up).

In conclusion we have shown 
\bp
\label{propM-JMU}
The (modified) Malgrange differential $\Omega$ restricted to the manifold of isomonodromic deformations coincides with the Jimbo-Miwa-Ueno differential.
\ep
\section{Applications and examples}
\label{Examples}
\subsection{Painlev\'e\ II equation}
We single out the second Painlev\'e\ for its relative simplicity in formul\ae\ and as an illustration of potential applications to other Painlev\'e\ equations. We follow the setup in \cite{Its:2002p9}. 
The Riemann--Hilbert description is depicted in the picture, with the notations and condition
\bea
L(s):= \le[\begin{array}{cc}
1 & 0 \\
s \,{\rm e}^{\frac {i4}3 z^3 + i t z} & 1
\end{array}\ri]\ ,\qquad U(s):= \le[\begin{array}{cc}
1 & s \,{\rm e}^{-\frac {i4}3 z^3 - i t z}\\
0 & 1
\end{array}\ri]\\
s_1-s_2+s_3 +s_1 s_2 s_3 =0\label{nomonodromyPII}
\eea
Condition \ref{nomonodromyPII} is the condition that the products of the jumps at the origin is the identity. The Riemann--Hilbert problem is then that of finding $\Gamma(z)$ uniformly bounded, with the indicated jumps and the normalization condition 
\be
\Gamma(z) \sim \1 + \mathcal O(z^{-1}) 
\ee
The matrix $T(z)$  and the Stokes' matrices are simply 
\be
T(z) =  - \le(\frac {4i}3 z^3 + it  z\ri)\sigma_3\ ,\qquad \sigma_3:= \le[\begin{array}{cc}
1 & 0\\
0& -1
\end{array}
\ri]
\ee

\begin{wrapfigure}{r}{0.4\textwidth}
\resizebox{0.4\textwidth}{!}{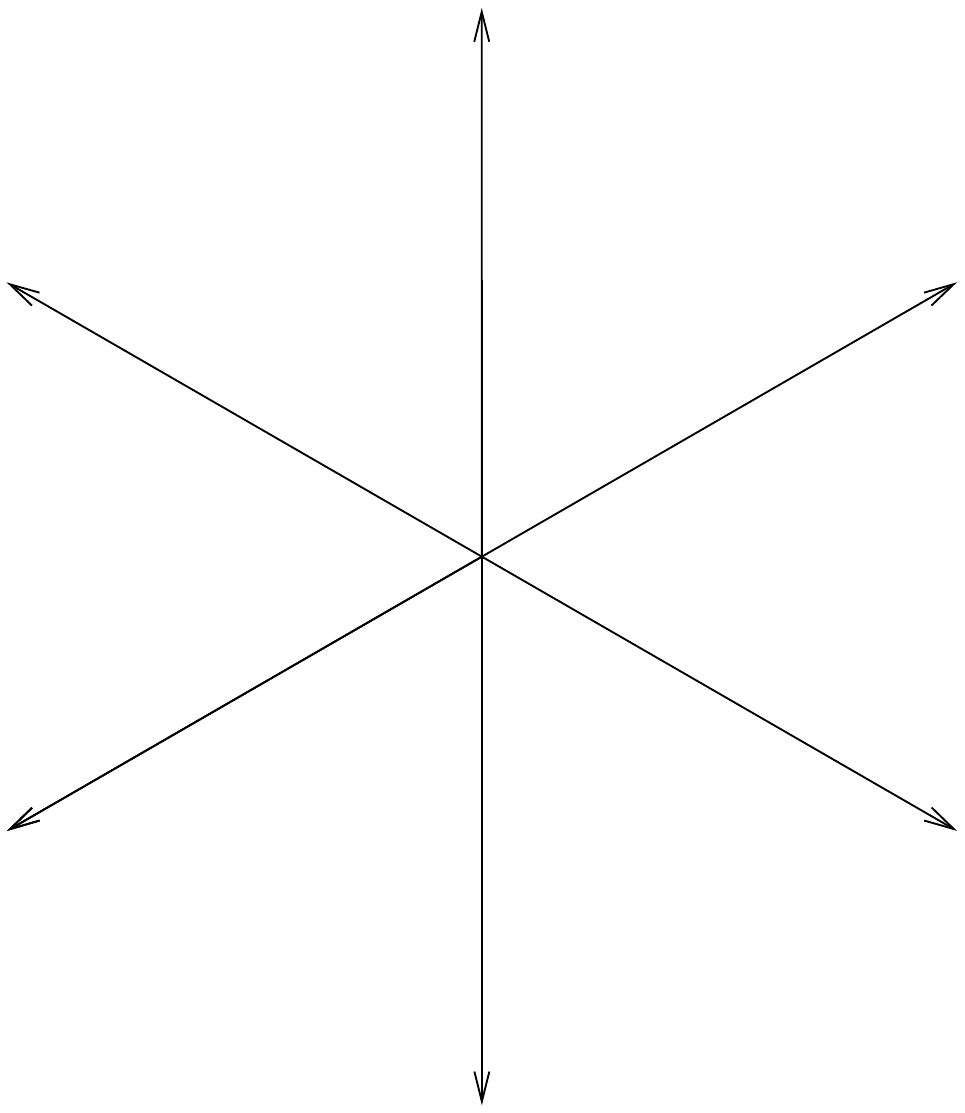}
\end{wrapfigure}

and the matrix $\Psi(z):= \Gamma(z) {\rm e}^{T}$ has constant jumps (on the anti-Stokes lines) and satisfies 
\bea
\Psi'(z)\Psi(z)^{-1} = -i\le(4 z^2 + t + 2 u^2\ri) \sigma_3 - 4 u  z \sigma_2 - 2 v \sigma_1=\cr=
\pmatrix{-4i & 0 \cr 0 & 4i} z^2 + \pmatrix{ 0 & 4iu\cr -4i u} z + \pmatrix{ -it-2iu^2 & -2v\cr -2v & it+2iu^2}  
\label{PIIODE}\\
\sigma_1:= \le[
\begin{array}{cc}
0 & 1\\ 1 & 0 
\end{array}
\ri]\ ,\ \ \ \sigma_2:= \le[
\begin{array}{cc}
0 & -i\\ i & 0 
\end{array}
\ri]\ ,\ \ v = q'(t)
\eea
Under an isomonodromic deformation the coefficients $u, v$ in (\ref{PIIODE}) must evolve as functions of $x$ and --in particular--
\be
u = u(t; \vec s) = 2\lim_{z\to \infty}  z \, \Gamma_{12}(z; t, \vec s)\label{PIIlim}
\ee
(the limit does not depend on the sector we choose)
solves the second Painlev\'e\ equation 
\be
\frac {\d^2 u }{\d t^2} = 2u^3 + t u
\ee
By direct computations we have 
\bea
\omega_{_{JMU}} = \le(\le(\frac {\d u}{\d t}\ri)^2 - t u^2 - u^4\ri) \d t = \pa_t \ln \tau(t;\vec s) \d t \ ,\ \ \ 
\pa_t^2 \ln \tau(t;\vec s) = u(t;\vec s)^2\ .
\eea
Generically we can solve the condition (\ref{nomonodromyPII}) for $s_3$ and use $s_1,s_2$ as independent variables: more appropriately one should consider the parameter space as the algebraic manifold specified by (\ref{nomonodromyPII}).
Introduce the matrix kernel 
\be
\KK(x,y):= \frac {\Gamma(x)^{-1} \Gamma(y)}{x-y}\ .
\ee
 We then have 
\bea
\pa_{s_1} \ln \tau(t;\vec s) &\&= 
-\int_{\ell_1} \tr \le(\Gamma^{-1} \Gamma' \le[\begin{array}{cc}
0 &0\\
{\rm e}^{\frac {i4}3 x^3 + i t x}&0
\end{array}\ri]\ri)\dd
 + \frac {1+s_2^2}{(1+s_1s_2)^2} \int_{\ell_3}  \tr \le(\Gamma^{-1} \Gamma' \le[\begin{array}{cc}
0 &{\rm e}^{-\frac {i4}3 x^3 - i t x}\cr
0&0
\end{array}\ri]\ri)\dd =\cr
&\&=
- \int_{\ell_1}\KK_{12}(x,x) {\rm e}^{\frac {i4}3 x^3 + i t x}\dd 
+\frac {1+s_2^2}{(1+s_1s_2)^2}  \int_{\ell_3}\KK_{21}(x,x) {\rm e}^{-\frac {i4}3 x^3 - i t x}\dd
\eea
where the boundary--value indication is irrelevant since the indicated matrix element does not have a jump on the corresponding line.
Using formula (\ref{secondder}) for the second derivatives and then (\ref{pakern}) one may derive an integral formula for the derivative $\pa_t^2 \pa_{s_1} \ln \tau$; however, in view of (\ref{PIIlim}) it is simpler to use the variational formula (\ref{pagamma}) directly, thus yielding 
\bea
\pa_{s_1} u(t;\vec s) = 2\lim_{z\to \infty} z\,\pa_{s_1} \Gamma_{12} =\cr=2\lim_{z\to\infty}  \int_{\ell_1} \frac {z\le(\Gamma(x) 
 \le[\begin{array}{cc}
0 &0\\
{\rm e}^{\frac {i4}3 x^3 + i t x}&0
\end{array}\ri]
 \Gamma^{-1}(x)\Gamma(z)\ri)_{12}}{x-z} \dd+\cr
  -2\lim_{z\to\infty}
 \frac {1+s_2^2}{(1+s_1s_2)^2}  \int_{\ell_3} \frac {z\le(\Gamma(x) 
 \le[\begin{array}{cc}
0 &{\rm e}^{-\frac {i4}3 x^3 - i t x}\\
0&0
\end{array}\ri]
 \Gamma^{-1}(x)\Gamma(z)\ri)_{12}}{x-z} \dd
\eea
Since $\Gamma(z)\to \1$ we have 
\be
\pa_{s_1} u(t;\vec s) =   2 \frac {1+s_2^2}{(1+s_1s_2)^2} \int_{\ell_3} \Gamma_{11}(x) ^2 {\rm e}^{-\frac {i4}3 x^3 - i t x}\dd - 2 \int_{\ell_1} \Gamma_{12}(x)^2
{\rm e}^{\frac {i4}3 x^3 + i t x}\dd 
\ee
\subsection{Variation of finite Toeplitz determinants for discontinuous symbols}
Let $S^1= \{ |z|=1\}$. Suppose that we have the following data:
\begin{itemize}
\item A finite partition of $S^1$ in  subarcs $\gamma_j$, with $\beta_j$ being the separating points;
\item a collection of  functions $a_j(z):\gamma_j\to \C$ which are analytic in a neighborhood of $\gamma_j$.
\end{itemize} 
Define $a(z):S^1\to \C$ as the piecewise analytic function that coincides with $a_j$ on each (interior of) $\gamma_j$.
The $n$-th Toeplitz determinant is defined as 
\be
T_n[a]:= \det[\mu_{j-i}]_{i,j=1\dots n}\ ,\qquad \mu_j:= \oint_{|z|=1} z^j a(z)\frac {\d z}{2i\pi z}\ ,\ \ T_0[a]:=1
\ee
It was shown in \cite{Bertola:MomentTau} that $T_n[a]$ (and more general objects) are isomonodromic tau functions for the following RHP;
\bea
\Gamma(z) = \Gamma_n(z)\ ,\ 
\Gamma_+ = \Gamma_- \le[\begin{array}{cc}
1 &w (z)\\
0 & 1
\end{array}
\ri]\cr 
\Gamma(z) = \Gamma_n(z) = \le(\1 + \mathcal O(z^{-1}) \ri) z^{n\sigma_3}\cr 
\Gamma(z) = \mathcal O(\ln (z-\beta_j))\label{RHPTop0}\ ,\ \ w(z):= z^{n-1} a(z)
\eea

In order to reduce to the general setting of Sect. \ref{RHPsec} we should add a big counterclockwise circle (the ``formal monodromy'' circle) with jump $z^{n\sigma_3}$ and replace the asymptotics with $\Gamma(\infty)=\1$; to take care of the logarithmic growth at the endpoints $\beta_j$ we 
take $\epsilon$--circles centered at the $\beta_j$ (with $\epsilon$ sufficiently small) and re-define $\Gamma$ as follows
\be
\wt \Gamma(z) := \Gamma(z) \le[\begin{array}{cc}
1 & -\mathcal C[w](z)\\
0 & 1 
\end{array}\ri], \ |z-\beta_j|<\epsilon\ ,\qquad \mathcal C[w](z):=  \oint_{|x|=1}\frac {w(x)}{x-z}\dd
\ee 

\begin{wrapfigure}{r}{0.4\textwidth}
\resizebox{0.4\textwidth}{!}{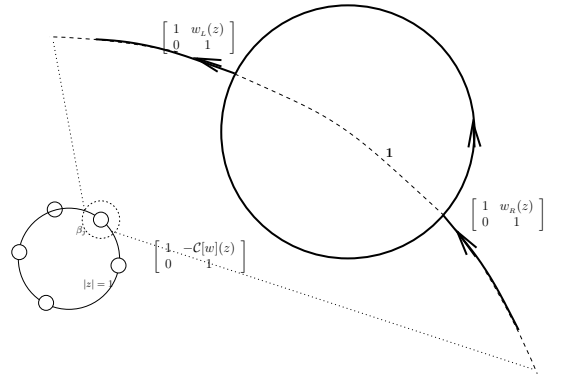}
\caption{The modification to the RHP to guarantee that the solution is bounded everywhere: here $\mathcal C[w](z):= \oint\frac {w(x)}{x-z}\dd$ and the new RHP has no jump within the small disk centered at $\beta_j$.} 
\label{figTop}
\vspace{-2cm}
\end{wrapfigure}

Thus we recast the problem (\ref{RHPTop0}) in the equivalent one
\bea
&\& \Gamma_+ = \Gamma_- \le[\begin{array}{cc}
1 & z^n a(z)\\
0 & 1
\end{array}
\ri]\ ,\ |z|=1,\ \min_j |z-\beta_j|>\epsilon\cr
&\& \Gamma_+ = \Gamma_-\le[\begin{array}{cc}
1 & -\mathcal C[w](z)\\
0 & 1 
\end{array}\ri] \ \ |x-\beta_j|=\epsilon \cr 
&\& \Gamma_+ = \Gamma_- z^{n\sigma_3}\ ,\ \ |z|=R>1 \cr
&\& \Gamma(z) = \Gamma_n(z) = \le(\1 + \mathcal O(z^{-1}) \ri)\label{RHPTop}
\eea
and $\Gamma(z)$ uniformly bounded  in $\C$.

 Note that the additional jump on $|z|=R$ (oriented counterclockwise) is independent of the symbol and hence undergoes no deformation. Therefore this has no impact in the definition of the differential  $\Omega_n(\bullet) = \Omega(\bullet; [\Gamma_n])$.

Note that the first column of the solution $\Gamma$ to problem (\ref{RHPTop}) consists of {\bf polynomials} (of degree $\leq n$) for $|z|<R$; indeed it has no jumps on $|z|=1$ and the added small circles (due to the triangularity of the jump-matrices). 
Outside the circle $|z|=R$, said column divided by $z^n$ is  bounded at infinity, which proves the assertion. 
 
Similarly, the second row of $\Gamma^{-1}$ (which solves a RHP with the jump matrix on the left) also consists of polynomials, by parallel arguments.

Thus the integrand in the definition of the differential $\Omega$ is in fact the integral of a polynomial kernel\footnote{A fact which is well known and key to matrix model computations.} 
\be
\KK_{21}(z,w):=\frac{\le[\Gamma^{-1}(z)\Gamma(w)\ri]_{21}}{z-w}
\ee
 of degree $\leq n-1$ in both variables (it has no jumps, it is regular on the diagonal $z=w$ and bounded by a power at infinity separately in each variable).

Now,  the computations in \cite{Bertola:MomentTau} where done for an {\em isomonodromic} tau function, however  the proof passes through without change to the extended $\Omega$-differential. Namely, the proof consisted in  showing that 
\be
\Omega_{n+1} -\Omega_n = \delta \ln \le(\frac {T_{n+1}[a]}{T_{n}[a]}\ri)
\ee
where $\Omega_n$ denotes the differential evaluated on the solution $\Gamma_{n}$ of (\ref{RHPTop}). Since the two RHPs differ by an elementary Schlesinger transformation of the type discussed in Sect. \ref{SecSchles}, this part is identical. Moreover, $\Omega_0\equiv 0$ because the integrands in (\ref{recurv}) and (\ref{omegaM}) are identically zero (all the matrices are upper triangular and $\Xi_\pa$ is strictly upper triangular),  which means that $\tau_0 = \exp \int \Omega_0$ is a constant independent of the symbol, and can be taken $1$. Since $T_0[a]=1$ as well, this implies that 
\be
\delta \ln T_n[a] = \Omega_n
\ee
where $\delta$ is the total differential w.r.t. any parameters may appear in the symbol and the positions of the endpoints.

\paragraph{Dependence on the $\beta_j$:} the dependence on each of the $\beta_j$ is only in the jumps around the corresponding circle, and we have 
\be
\Xi_{\pa_{\beta_j}} = \le[
\begin{array}{cc}
0 &\frac{-\Delta_j(\vec \beta)}{2i\pi} \frac 1{z-\beta_j}\\
0 & 0 
\end{array}
\ri]\ ,\qquad \Delta_j(\vec \beta):=  { w_{_R}(\beta_j) - w_{_L}(\beta_j)} 
\ee
where the subscripts $_{_{L,R}}$ indicate the boundary value on the Left/Right of $w$ along the contour, relative to the orientation of the contour, namely,  $\Delta$ is the jump of the symbol.
This shows --a well--known fact-- that 
\be
\pa_{\beta_j} T_n[a] = -\oint_{|x-\beta_j|=\epsilon}\hspace{-20pt} \KK_{21}(x,x) \frac{-\Delta_j(\vec \beta)}{2i\pi} \frac 1{x-\beta_j}\dd =\frac {\Delta_j(\vec \beta)}{2i\pi} \KK_{21}(\beta_j,\beta_j)
\ee
\begin{wrapfigure}{r}{0.3\textwidth}
\resizebox{0.3\textwidth}{!}{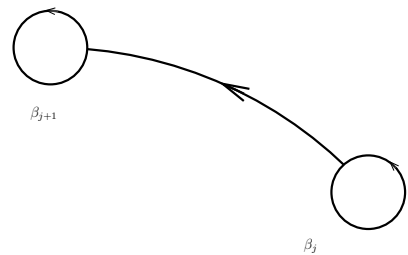}
\caption{The ``dumbbell'' contour}
\label{dumbbell}
\end{wrapfigure}

\paragraph{Dependence on the symbol.} Let $\pa$ denote a deformation of one of the functions $a_j$; then the jump matrices $M(z)$ deforms only on the arc $\gamma_j$ and the small circles we have added, on a ``dumb-bell'' contour (Fig. \ref{dumbbell}). In particular, on the two circles the integral for $\Omega$ reduces to  
\bea
\oint_{|z-\beta_j|=\epsilon} \KK_{21}(x,x) \mathcal C[\pa w](x)\dd\\
\mathcal C[\pa w](z) =  \int_{\gamma_j} \frac {x^n\pa a_j(x;\vec s)}{x-z} \dd
\eea
Since, $\KK_{21}(x,x)$ is a analytic in the interior of the $\epsilon$--circle (in fact, it is a polynomial!) then we can perform the integration by ``contour deformation'' leaving an integral of $\KK_{21}(x,x)$ against the jump of $\mathcal C[\pa w]$. In conclusion, taking into consideration all the contours we have simply 
\be
\pa T_n[a]= \int_{\gamma_j} \KK_{21}(x,x) x^n \pa a_j(x;\vec s) \dd
\ee

In general we have simply 
\bea
\pa \ln T_n[a] = \oint_{|z|=1} \tr \le(\Gamma^{-1}(x)\Gamma'(x)\le[\begin{array}{cc}
0 &x^n \pa_s a(x;\vec s) \\
0  & 0 
\end{array}\ri]\ri)\dd =\\= \oint \KK_{21}(x,x) x^n \pa_s a(x;\vec s)\dd\\
\KK(x,y):= \frac {\Gamma^{-1}(x) \Gamma(y)}{x-y}
\eea
\bx
A particular case is if $a_j(x;s) = s_j$ are constants, in which case the above reduces to an integral on the sub-arc $\gamma_j$
\be
\pa_{s_j} \ln T_n[a] = \int_{\gamma_j}  \KK_{21}(x,x) x^n\dd
\ee
\ex
\subsection{Hankel  and shifted T\"oplitz determinants}
The notion of {\em semiclassical symbols} was defined in \cite{Marcellan:Semiclassical1} and it was shown in \cite{Bertola:MomentTau} that the corresponding Hankel/Toeplitz determinant were isomonodromic tau function, to within an explicit non-vanishing factor.

The weights we are considering are all of the {\em semiclassical type} as defined in \cite{Magnus:Painleve, Marcellan:Semiclassical1, Marcellan:Semiclassical2, Bertola:Semiiso, Bertola:BilinearJAT}. This  means that they are of the form $\mu(x) = {\rm e}^{-V(x)}$ with $V'(x)$ an arbitrary rational function.

They are integrated over contours $\gamma_j$ which can be arbitrary contours in the complex plane as long as  all integrals $\int_{\gamma_j} x^k \mu(x) \d x $ are convergent integrals. The range for $k$ will be either $\N$ or $\Z$, depending on the situation; a detailed description of the contours can be found in \cite{Bertola:Semiiso, Bertola:LOPs} and we refer thereto for a more detailed discussion.

We will choose arbitrary complex constants $\varkappa_j$ for each contour $\gamma_j$ and use the notation 
\be
\int_\kappa x^k \mu(x) \d x := \sum \varkappa_j \int _{\gamma_j} x^k \mu(x) \d x  = \mu_k
\ee
We will also use the notation $\kappa: \C \to \C$ to indicate  the locally constant function that takes the (constant) value $\varkappa_j$ on the corresponding contour $\gamma_j$.  The Markov function (sometimes referred to as Weyl function) for these semiclassical weights is simply defined as the locally analytic function on $\C \setminus \cup \gamma_j$ given by 
\be
W(x):= \int_\kappa \frac{ \mu(\zeta) \d \zeta}{\zeta -x}\  ,\qquad
\kappa := \sum_j\varkappa_j  \chi_{\gamma_j}(x)
\ee
The function $W(x)$ has logarithmic growth at the {\bf hard--edges} i.e. the  endpoints of contours $\gamma_j$ where $\mu$ is $\mathcal O(1)$. In this case one verifies that $W(x) = \mathcal O(\ln |x-a|)$, where $a$ is the hard--edge point.

Consider the jump matrix supported on the contours $\bigcup \gamma_j$
\be
M(z):= \le[\begin{array}{cc}
1 & \varkappa_j {\rm e}^{-V(z)}\\
0 & 1 
\end{array}\ri]\ ,\ \ z\in \gamma_j.
\ee
Consider the following Riemann--Hilbert problems for a matrix $\Gamma = \Gamma^{(n,\ell)}$ parametrized by two integers $n,\ell$
\bea
\Gamma_+ &\&= \Gamma_- M\cr
\Gamma(z) &\& = \le\{
\begin{array}{cc}
\mathcal O(1) \le[
\begin{array}{cc}
1 & 0 \\ 
0& z^{n-\ell - 1}
\end{array}
\ri] &  z\to 0\\
(\1 + \mathcal O(z^{-1})) \le[
\begin{array}{cc}
z^n & 0 \\ 
0& z^{-\ell - 1}
\end{array}
\ri] &  z\to \infty
\end{array}\ri. 
\eea
and  $\Gamma(z) = \mathcal O(\ln (z-a))$ at any ``hard-edge''.

By definition of Schlesinger transformations (Sect. \ref{SecSchles}) we see that all these problems are successive Schlesinger transformations of the problem $n=0, \ell=1$, which has the immediate solution 
\be
\Gamma^{(0,1)}(z) = \le[
\begin{array}{cc}
1 & \int_{\varkappa} \frac { {\rm e}^{-V(x)}}{x-z} \dd\\
0 &1
\end{array}
\ri]
\ee
It is also apparent that the differential $\omega_{_M}$ (or $\Omega$, they coincide here because the term $\vartheta$ in Def. \ref{defOmega} vanishes identically, being the jumps $\1 + $upper triangular) are closed differentials in the deformations of $V$, the endpoints and in the $\varkappa_j$'s.

Moreover, comparing the change $\omega_{_M} (\pa;[\Gamma^{(n,\ell)}])$  between two problems with $n\mapsto n\pm 1$ or $\ell\mapsto \ell \pm 1$ one sees exactly as in \cite{Bertola:MomentTau} that 
\be
\pa \ln \Delta_n^\ell  = \omega_{_M}(\pa; [\Gamma^{(n,\ell)}])
\ee
where $\Delta_n^\ell$ are the shifter Toeplitz determinants
\bea
\Delta_n^\ell :=\det\pmatrix {\mu_{\ell} & \mu_{\ell+1} & \cdots &\mu_{\ell+n-1}\cr
\mu_{\ell-1} & \mu_\ell & \cdots & \mu_{\ell+n-2}\cr
 & \ddots & \ddots & \cr
\mu_{\ell-n+1} & \mu_{\ell-n+2} & \cdots & \mu_{\ell}}
\ \ 
\Delta_0^\ell\equiv 1 \ ,\ \ \Delta_{-n}^{\ell} \equiv 0\nonumber 
\eea
Here the case of Hankel determinants corresponds to $\ell = n-1$; then $\Delta_{n}^{n-1}$ is --up to a permutation of columns, hence a sign-- a Hankel determinant.

In particular the derivatives w.r.t. the parameters $\varkappa_j$ are 
\bea
\pa_{\varkappa_j} \ln \Delta_n^\ell  =\int_{\gamma_j} K^{(n,\ell)}(x,x) {\rm e}^{-V(x)} \dd\\
K^{(n,\ell)} (x,y) =\frac{ \le[\Gamma^{(n,\ell)}(x)^{-1} \Gamma^{(n,\ell)}(y)\ri]_{21}}{x-y}\ .
\eea
This kernel is the usual Christoffel--Darboux kernel: for example, if $\ell=n-1$ (Hankel determinants) then 
\be
K^{(n,n-1)} (x,y)  = \frac 1{h_{n-1}} \frac {p_n(x)p_{n-1}(y) - p_{n-1}(x) p_n(y)}{x-y} = \sum_{j=0}^{n-1} \frac 1{h_j} p_j(x)p_j(y)
\ee
where $p_j$ are the monic orthogonal polynomials of exact degree $j$ for the moment-functional, namely, 
\be
\int_{\varkappa} p_j(x)p_k(x) {\rm e}^{-V(x)} \d x = h_j\delta_{jk}
\ee
\br
Note that the RHP for $\Gamma^{(n,\ell)}$ can be converted to a RHP with constant jumps for $\Psi := \Gamma {\rm e}^{-\frac V2 \sigma_3}$ and hence reduced to a rational ODE $\Psi' \Psi^{-1} = D(z)$. The resulting {\em isomonodromic} tau function is not exactly equal to  $\Delta_n^{\ell}$ because the RHP for $\Gamma^{(n,\ell)}$ and the RHP in the canonical form described in Sec. \ref{SecBirk} differ by a gauge in the sense of Sect. \ref{secGauge}. However the corresponding factor is easy to compute and this was accomplished in \cite{Bertola:MomentTau}.
\er
We conclude with the remark that the variational formul\ae\  for $\Delta_{n}^\ell$ are valid for very general weights, not necessarily of semiclassical type. Indeed we are really using only the results of Sect. \ref{RHPsec}, where the connection to rational ODE is not necessary.

\bibliographystyle{plain}
\bibliography{/Users/bertola/Documents/Papers/BibDeskLibrary}

\end{document}

%% file: HidingSing.pdf_t
\begin{picture}(0,0)%
\includegraphics{HidingSing.pdf}%
\end{picture}%
\setlength{\unitlength}{3947sp}%
\begingroup\makeatletter\ifx\SetFigFont\undefined%
\gdef\SetFigFont#1#2#3#4#5{%
  \reset@font\fontsize{#1}{#2pt}%
  \fontfamily{#3}\fontseries{#4}\fontshape{#5}%
  \selectfont}%
\fi\endgroup%
\begin{picture}(9077,6173)(3075,-4037)
\put(10351,314){\rotatebox{270.0}{\makebox(0,0)[lb]{\smash{{\SetFigFont{12}{14.4}{\familydefault}{\mddefault}{\updefault}{\color[rgb]{0,0,0}$C^{-1} {\rm e}^{2i\pi L} C$}%
}}}}}
\put(5251,239){\rotatebox{270.0}{\makebox(0,0)[lb]{\smash{{\SetFigFont{12}{14.4}{\familydefault}{\mddefault}{\updefault}{\color[rgb]{0,0,0}$M=C^{-1} {\rm e}^{2i\pi L} C$}%
}}}}}
\put(5026,-2836){\makebox(0,0)[lb]{\smash{{\SetFigFont{12}{14.4}{\familydefault}{\mddefault}{\updefault}{\color[rgb]{0,0,0}$a$}%
}}}}
\put(3601,-3361){\makebox(0,0)[lb]{\smash{{\SetFigFont{12}{14.4}{\familydefault}{\mddefault}{\updefault}{\color[rgb]{0,0,0}$\Gamma \sim Y(z)(z-a)^L C$}%
}}}}
\put(10201,-2761){\makebox(0,0)[lb]{\smash{{\SetFigFont{12}{14.4}{\familydefault}{\mddefault}{\updefault}{\color[rgb]{0,0,0}$a$}%
}}}}
\put(9001,-1036){\makebox(0,0)[lb]{\smash{{\SetFigFont{12}{14.4}{\familydefault}{\mddefault}{\updefault}{\color[rgb]{0,0,0}$\wh \Gamma=\Gamma$}%
}}}}
\put(9151,-3061){\makebox(0,0)[lb]{\smash{{\SetFigFont{12}{14.4}{\familydefault}{\mddefault}{\updefault}{\color[rgb]{0,0,0}$\wh \Gamma=\Gamma C^{-1} (z-a)^{-L} = \mathcal O(1)$}%
}}}}
\put(8701,-1936){\rotatebox{270.0}{\makebox(0,0)[lb]{\smash{{\SetFigFont{12}{14.4}{\familydefault}{\mddefault}{\updefault}{\color[rgb]{0,0,0}$C^{-1}(z-a)^{-L}$}%
}}}}}
\end{picture}%

%% file: Goodpoint.pdf_t
\begin{picture}(0,0)%
\includegraphics{Goodpoint.pdf}%
\end{picture}%
\setlength{\unitlength}{3947sp}%
\begingroup\makeatletter\ifx\SetFigFont\undefined%
\gdef\SetFigFont#1#2#3#4#5{%
  \reset@font\fontsize{#1}{#2pt}%
  \fontfamily{#3}\fontseries{#4}\fontshape{#5}%
  \selectfont}%
\fi\endgroup%
\begin{picture}(3730,2940)(4158,-5464)
\put(5926,-4561){\makebox(0,0)[lb]{\smash{{\SetFigFont{12}{14.4}{\familydefault}{\mddefault}{\updefault}{\color[rgb]{0,0,0}$\xi$}%
}}}}
\put(4520,-5386){\makebox(0,0)[lb]{\smash{{\SetFigFont{12}{14.4}{\familydefault}{\mddefault}{\updefault}{\color[rgb]{0,0,0}$\gamma_4$}%
}}}}
\put(7473,-2906){\makebox(0,0)[lb]{\smash{{\SetFigFont{12}{14.4}{\familydefault}{\mddefault}{\updefault}{\color[rgb]{0,0,0}$\gamma_2$}%
}}}}
\put(4173,-3106){\makebox(0,0)[lb]{\smash{{\SetFigFont{12}{14.4}{\familydefault}{\mddefault}{\updefault}{\color[rgb]{0,0,0}$\gamma_3$}%
}}}}
\put(7673,-4586){\makebox(0,0)[lb]{\smash{{\SetFigFont{12}{14.4}{\familydefault}{\mddefault}{\updefault}{\color[rgb]{0,0,0}$\gamma_1$}%
}}}}
\end{picture}%

%% file: RHP.pdf_t
\begin{picture}(0,0)%
\includegraphics{RHP.pdf}%
\end{picture}%
\setlength{\unitlength}{3947sp}%
\begingroup\makeatletter\ifx\SetFigFont\undefined%
\gdef\SetFigFont#1#2#3#4#5{%
  \reset@font\fontsize{#1}{#2pt}%
  \fontfamily{#3}\fontseries{#4}\fontshape{#5}%
  \selectfont}%
\fi\endgroup%
\begin{picture}(10736,5877)(884,-6805)
\put(8401,-3586){\makebox(0,0)[lb]{\smash{{\SetFigFont{12}{14.4}{\familydefault}{\mddefault}{\updefault}{\color[rgb]{0,0,0}${\rm e}^{2i\pi L_2}$}%
}}}}
\put(5776,-6586){\makebox(0,0)[lb]{\smash{{\SetFigFont{12}{14.4}{\familydefault}{\mddefault}{\updefault}{\color[rgb]{0,0,0}connection circle}%
}}}}
\put(2225,-5203){\makebox(0,0)[lb]{\smash{{\SetFigFont{12}{14.4}{\familydefault}{\mddefault}{\updefault}{\color[rgb]{0,0,0}$(z-a_1)^{-L_1}$}%
}}}}
\put(4651,-4561){\makebox(0,0)[lb]{\smash{{\SetFigFont{12}{14.4}{\familydefault}{\mddefault}{\updefault}{\color[rgb]{0,0,0}$C_1^{-1}$}%
}}}}
\put(8776,-1936){\makebox(0,0)[lb]{\smash{{\SetFigFont{12}{14.4}{\familydefault}{\mddefault}{\updefault}{\color[rgb]{0,0,0}$(z-a_2)^{-L_2}$}%
}}}}
\put(8776,-1336){\makebox(0,0)[lb]{\smash{{\SetFigFont{12}{14.4}{\familydefault}{\mddefault}{\updefault}{\color[rgb]{0,0,0}$C_2^{-1}$}%
}}}}
\put(9676,-3211){\makebox(0,0)[lb]{\smash{{\SetFigFont{12}{14.4}{\familydefault}{\mddefault}{\updefault}{\color[rgb]{0,0,0}${\rm e}^{-T(z)} $}%
}}}}
\put(7651,-5086){\makebox(0,0)[lb]{\smash{{\SetFigFont{12}{14.4}{\familydefault}{\mddefault}{\updefault}{\color[rgb]{0,0,0}$(z-a_2)^{L_2}{\rm e}^{T(z)}  S_\nu {\rm e}^{-T(z)}(z-a_2)^{-L_2}$}%
}}}}
\put(3500,-3328){\makebox(0,0)[lb]{\smash{{\SetFigFont{12}{14.4}{\familydefault}{\mddefault}{\updefault}{\color[rgb]{0,0,0}${\rm e}^{2i\pi L_1}$}%
}}}}
\put(3376,-4111){\makebox(0,0)[lb]{\smash{{\SetFigFont{12}{14.4}{\familydefault}{\mddefault}{\updefault}{\color[rgb]{0,0,0}+}%
}}}}
\put(3755,-1933){\makebox(0,0)[lb]{\smash{{\SetFigFont{12}{14.4}{\familydefault}{\mddefault}{\updefault}{\color[rgb]{0,0,0}$C_1^{-1}{\rm e}^{2i\pi L_1} C_1$}%
}}}}
\put(4726,-2311){\makebox(0,0)[lb]{\smash{{\SetFigFont{12}{14.4}{\familydefault}{\mddefault}{\updefault}{\color[rgb]{0,0,0}{\bf +}}%
}}}}
\put(4126,-4786){\makebox(0,0)[lb]{\smash{{\SetFigFont{12}{14.4}{\familydefault}{\mddefault}{\updefault}{\color[rgb]{0,0,0}+}%
}}}}
\put(4351,-2236){\makebox(0,0)[lb]{\smash{{\SetFigFont{12}{14.4}{\familydefault}{\mddefault}{\updefault}{\color[rgb]{0,0,0}--}%
}}}}
\put(3676,-4112){\makebox(0,0)[lb]{\smash{{\SetFigFont{12}{14.4}{\familydefault}{\mddefault}{\updefault}{\color[rgb]{0,0,0}--}%
}}}}
\put(4391,-4953){\makebox(0,0)[lb]{\smash{{\SetFigFont{12}{14.4}{\familydefault}{\mddefault}{\updefault}{\color[rgb]{0,0,0}--}%
}}}}
\put(6727,-2083){\makebox(0,0)[lb]{\smash{{\SetFigFont{12}{14.4}{\familydefault}{\mddefault}{\updefault}{\color[rgb]{0,0,0}+}%
}}}}
\put(6513,-2290){\makebox(0,0)[lb]{\smash{{\SetFigFont{12}{14.4}{\familydefault}{\mddefault}{\updefault}{\color[rgb]{0,0,0}--}%
}}}}
\put(7351,-2311){\makebox(0,0)[lb]{\smash{{\SetFigFont{12}{14.4}{\familydefault}{\mddefault}{\updefault}{\color[rgb]{0,0,0}${\rm e}^{2i\pi L_2}$}%
}}}}
\put(6151,-1486){\makebox(0,0)[lb]{\smash{{\SetFigFont{12}{14.4}{\familydefault}{\mddefault}{\updefault}{\color[rgb]{0,0,0}$C_2^{-1}{\rm e}^{2i\pi L_2} C_2$}%
}}}}
\put(10565,-4819){\makebox(0,0)[lb]{\smash{{\SetFigFont{12}{14.4}{\familydefault}{\mddefault}{\updefault}{\color[rgb]{0,0,0}--}%
}}}}
\put(10351,-4772){\makebox(0,0)[lb]{\smash{{\SetFigFont{12}{14.4}{\familydefault}{\mddefault}{\updefault}{\color[rgb]{0,0,0}+}%
}}}}
\put(10858,-4412){\makebox(0,0)[lb]{\smash{{\SetFigFont{12}{14.4}{\familydefault}{\mddefault}{\updefault}{\color[rgb]{0,0,0}+}%
}}}}
\put(11111,-4406){\makebox(0,0)[lb]{\smash{{\SetFigFont{12}{14.4}{\familydefault}{\mddefault}{\updefault}{\color[rgb]{0,0,0}--}%
}}}}
\put(6826,-2911){\makebox(0,0)[lb]{\smash{{\SetFigFont{12}{14.4}{\familydefault}{\mddefault}{\updefault}{\color[rgb]{0,0,0}$\beta_2$}%
}}}}
\put(8571,-5993){\makebox(0,0)[lb]{\smash{{\SetFigFont{11}{13.2}{\familydefault}{\mddefault}{\updefault}{\color[rgb]{0,0,0}toral circle}%
}}}}
\put(3032,-3060){\makebox(0,0)[lb]{\smash{{\SetFigFont{11}{13.2}{\familydefault}{\mddefault}{\updefault}{\color[rgb]{0,0,0}$\beta_1$}%
}}}}
\put(8085,-6441){\makebox(0,0)[lb]{\smash{{\SetFigFont{11}{13.2}{\familydefault}{\mddefault}{\updefault}{\color[rgb]{0,0,0}formal monodromy circle}%
}}}}
\end{picture}%

%% file: RHPPsi.pdf_t
\begin{picture}(0,0)%
\includegraphics{RHPPsi.pdf}%
\end{picture}%
\setlength{\unitlength}{3947sp}%
\begingroup\makeatletter\ifx\SetFigFont\undefined%
\gdef\SetFigFont#1#2#3#4#5{%
  \reset@font\fontsize{#1}{#2pt}%
  \fontfamily{#3}\fontseries{#4}\fontshape{#5}%
  \selectfont}%
\fi\endgroup%
\begin{picture}(10736,5877)(884,-6805)
\put(6151,-1486){\makebox(0,0)[lb]{\smash{{\SetFigFont{12}{14.4}{\familydefault}{\mddefault}{\updefault}{\color[rgb]{0,0,0}$C_2^{-1}{\rm e}^{2i\pi L_2} C_2$}%
}}}}
\put(10858,-4412){\makebox(0,0)[lb]{\smash{{\SetFigFont{12}{14.4}{\familydefault}{\mddefault}{\updefault}{\color[rgb]{0,0,0}+}%
}}}}
\put(11111,-4406){\makebox(0,0)[lb]{\smash{{\SetFigFont{12}{14.4}{\familydefault}{\mddefault}{\updefault}{\color[rgb]{0,0,0}--}%
}}}}
\put(6826,-2911){\makebox(0,0)[lb]{\smash{{\SetFigFont{12}{14.4}{\familydefault}{\mddefault}{\updefault}{\color[rgb]{0,0,0}$\beta_2$}%
}}}}
\put(5776,-6586){\makebox(0,0)[lb]{\smash{{\SetFigFont{12}{14.4}{\familydefault}{\mddefault}{\updefault}{\color[rgb]{0,0,0}connection circle}%
}}}}
\put(4651,-4561){\makebox(0,0)[lb]{\smash{{\SetFigFont{12}{14.4}{\familydefault}{\mddefault}{\updefault}{\color[rgb]{0,0,0}$C_1^{-1}$}%
}}}}
\put(8776,-1336){\makebox(0,0)[lb]{\smash{{\SetFigFont{12}{14.4}{\familydefault}{\mddefault}{\updefault}{\color[rgb]{0,0,0}$C_2^{-1}$}%
}}}}
\put(7651,-5086){\makebox(0,0)[lb]{\smash{{\SetFigFont{12}{14.4}{\familydefault}{\mddefault}{\updefault}{\color[rgb]{0,0,0}$ S_\nu$}%
}}}}
\put(3500,-3328){\makebox(0,0)[lb]{\smash{{\SetFigFont{12}{14.4}{\familydefault}{\mddefault}{\updefault}{\color[rgb]{0,0,0}${\rm e}^{2i\pi L_1}$}%
}}}}
\put(3376,-4111){\makebox(0,0)[lb]{\smash{{\SetFigFont{12}{14.4}{\familydefault}{\mddefault}{\updefault}{\color[rgb]{0,0,0}+}%
}}}}
\put(3755,-1933){\makebox(0,0)[lb]{\smash{{\SetFigFont{12}{14.4}{\familydefault}{\mddefault}{\updefault}{\color[rgb]{0,0,0}$C_1^{-1}{\rm e}^{2i\pi L_1} C_1$}%
}}}}
\put(4726,-2311){\makebox(0,0)[lb]{\smash{{\SetFigFont{12}{14.4}{\familydefault}{\mddefault}{\updefault}{\color[rgb]{0,0,0}{\bf +}}%
}}}}
\put(4126,-4786){\makebox(0,0)[lb]{\smash{{\SetFigFont{12}{14.4}{\familydefault}{\mddefault}{\updefault}{\color[rgb]{0,0,0}+}%
}}}}
\put(4351,-2236){\makebox(0,0)[lb]{\smash{{\SetFigFont{12}{14.4}{\familydefault}{\mddefault}{\updefault}{\color[rgb]{0,0,0}--}%
}}}}
\put(3676,-4112){\makebox(0,0)[lb]{\smash{{\SetFigFont{12}{14.4}{\familydefault}{\mddefault}{\updefault}{\color[rgb]{0,0,0}--}%
}}}}
\put(4391,-4953){\makebox(0,0)[lb]{\smash{{\SetFigFont{12}{14.4}{\familydefault}{\mddefault}{\updefault}{\color[rgb]{0,0,0}--}%
}}}}
\put(6727,-2083){\makebox(0,0)[lb]{\smash{{\SetFigFont{12}{14.4}{\familydefault}{\mddefault}{\updefault}{\color[rgb]{0,0,0}+}%
}}}}
\put(6513,-2290){\makebox(0,0)[lb]{\smash{{\SetFigFont{12}{14.4}{\familydefault}{\mddefault}{\updefault}{\color[rgb]{0,0,0}--}%
}}}}
\put(7351,-2311){\makebox(0,0)[lb]{\smash{{\SetFigFont{12}{14.4}{\familydefault}{\mddefault}{\updefault}{\color[rgb]{0,0,0}${\rm e}^{2i\pi L_2}$}%
}}}}
\put(10565,-4819){\makebox(0,0)[lb]{\smash{{\SetFigFont{12}{14.4}{\familydefault}{\mddefault}{\updefault}{\color[rgb]{0,0,0}--}%
}}}}
\put(10351,-4772){\makebox(0,0)[lb]{\smash{{\SetFigFont{12}{14.4}{\familydefault}{\mddefault}{\updefault}{\color[rgb]{0,0,0}+}%
}}}}
\put(8085,-6441){\makebox(0,0)[lb]{\smash{{\SetFigFont{11}{13.2}{\familydefault}{\mddefault}{\updefault}{\color[rgb]{0,0,0}formal monodromy circle}%
}}}}
\put(8571,-5993){\makebox(0,0)[lb]{\smash{{\SetFigFont{11}{13.2}{\familydefault}{\mddefault}{\updefault}{\color[rgb]{0,0,0}toral circle}%
}}}}
\put(3032,-3060){\makebox(0,0)[lb]{\smash{{\SetFigFont{11}{13.2}{\familydefault}{\mddefault}{\updefault}{\color[rgb]{0,0,0}$\beta_1$}%
}}}}
\end{picture}%

%% file: JMU.pdf_t
\begin{picture}(0,0)%
\includegraphics{JMU.pdf}%
\end{picture}%
\setlength{\unitlength}{3947sp}%
\begingroup\makeatletter\ifx\SetFigFont\undefined%
\gdef\SetFigFont#1#2#3#4#5{%
  \reset@font\fontsize{#1}{#2pt}%
  \fontfamily{#3}\fontseries{#4}\fontshape{#5}%
  \selectfont}%
\fi\endgroup%
\begin{picture}(7948,7880)(710,-6626)
\put(7588,-3955){\makebox(0,0)[lb]{\smash{{\SetFigFont{20}{24.0}{\familydefault}{\mddefault}{\updefault}{\color[rgb]{0,0,0}--}%
}}}}
\put(7204,-3870){\makebox(0,0)[lb]{\smash{{\SetFigFont{20}{24.0}{\familydefault}{\mddefault}{\updefault}{\color[rgb]{0,0,0}+}%
}}}}
\put(4710,-2652){\makebox(0,0)[lb]{\smash{{\SetFigFont{12}{14.4}{\familydefault}{\mddefault}{\updefault}{\color[rgb]{0,0,0}$\epsilon$}%
}}}}
\put(4057,-5285){\makebox(0,0)[lb]{\smash{{\SetFigFont{17}{20.4}{\familydefault}{\mddefault}{\updefault}{\color[rgb]{0,0,0}toral circle}%
}}}}
\put(3089,589){\makebox(0,0)[lb]{\smash{{\SetFigFont{17}{20.4}{\familydefault}{\mddefault}{\updefault}{\color[rgb]{0,0,0}formal monodromy circle}%
}}}}
\end{picture}%

%% file: PII.pdf_t
\begin{picture}(0,0)%
\includegraphics{PII.pdf}%
\end{picture}%
\setlength{\unitlength}{3947sp}%
\begingroup\makeatletter\ifx\SetFigFont\undefined%
\gdef\SetFigFont#1#2#3#4#5{%
  \reset@font\fontsize{#1}{#2pt}%
  \fontfamily{#3}\fontseries{#4}\fontshape{#5}%
  \selectfont}%
\fi\endgroup%
\begin{picture}(7710,8894)(2146,-8408)
\put(4351,-2911){\makebox(0,0)[lb]{\smash{{\SetFigFont{17}{20.4}{\familydefault}{\mddefault}{\updefault}{\color[rgb]{0,0,0}$L(s_3)$}%
}}}}
\put(7201,-2911){\makebox(0,0)[lb]{\smash{{\SetFigFont{17}{20.4}{\familydefault}{\mddefault}{\updefault}{\color[rgb]{0,0,0}$L(s_1)$}%
}}}}
\put(6076,-1261){\makebox(0,0)[lb]{\smash{{\SetFigFont{17}{20.4}{\familydefault}{\mddefault}{\updefault}{\color[rgb]{0,0,0}$U(s_2)$}%
}}}}
\put(3151,-5386){\makebox(0,0)[lb]{\smash{{\SetFigFont{17}{20.4}{\familydefault}{\mddefault}{\updefault}{\color[rgb]{0,0,0}$U(-s_1)$}%
}}}}
\put(6076,-6661){\makebox(0,0)[lb]{\smash{{\SetFigFont{17}{20.4}{\familydefault}{\mddefault}{\updefault}{\color[rgb]{0,0,0}$L(-s_2)$}%
}}}}
\put(8026,-5011){\makebox(0,0)[lb]{\smash{{\SetFigFont{17}{20.4}{\familydefault}{\mddefault}{\updefault}{\color[rgb]{0,0,0}$U(-s_3)$}%
}}}}
\put(8926,-2011){\makebox(0,0)[lb]{\smash{{\SetFigFont{17}{20.4}{\familydefault}{\mddefault}{\updefault}{\color[rgb]{0,0,0}$\ell_1$}%
}}}}
\put(6151,-211){\makebox(0,0)[lb]{\smash{{\SetFigFont{17}{20.4}{\familydefault}{\mddefault}{\updefault}{\color[rgb]{0,0,0}$\ell_2$}%
}}}}
\put(2701,-1936){\makebox(0,0)[lb]{\smash{{\SetFigFont{17}{20.4}{\familydefault}{\mddefault}{\updefault}{\color[rgb]{0,0,0}$\ell_3$}%
}}}}
\put(2551,-6361){\makebox(0,0)[lb]{\smash{{\SetFigFont{17}{20.4}{\familydefault}{\mddefault}{\updefault}{\color[rgb]{0,0,0}$\ell_4$}%
}}}}
\put(6076,-7936){\makebox(0,0)[lb]{\smash{{\SetFigFont{17}{20.4}{\familydefault}{\mddefault}{\updefault}{\color[rgb]{0,0,0}$\ell_5$}%
}}}}
\put(8851,-6136){\makebox(0,0)[lb]{\smash{{\SetFigFont{17}{20.4}{\familydefault}{\mddefault}{\updefault}{\color[rgb]{0,0,0}$\ell_6$}%
}}}}
\end{picture}%

%% file: Toplitz.pdf_t
\begin{picture}(0,0)%
\includegraphics{Toplitz.pdf}%
\end{picture}%
\setlength{\unitlength}{3947sp}%
\begingroup\makeatletter\ifx\SetFigFont\undefined%
\gdef\SetFigFont#1#2#3#4#5{%
  \reset@font\fontsize{#1}{#2pt}%
  \fontfamily{#3}\fontseries{#4}\fontshape{#5}%
  \selectfont}%
\fi\endgroup%
\begin{picture}(7800,5491)(2064,-2754)
\put(4156,-1066){\makebox(0,0)[lb]{\smash{{\SetFigFont{12}{14.4}{\familydefault}{\mddefault}{\updefault}{\color[rgb]{0,0,0}$\le[\begin{array}{cc} 1 &- \mathcal C[w](z) \\ 0 & 1\end{array}\ri]$}%
}}}}
\put(8851,-436){\makebox(0,0)[lb]{\smash{{\SetFigFont{12}{14.4}{\familydefault}{\mddefault}{\updefault}{\color[rgb]{0,0,0}$\le[\begin{array}{cc} 1 &w_{_R}(z) \\ 0 & 1\end{array}\ri]$}%
}}}}
\put(4276,2189){\makebox(0,0)[lb]{\smash{{\SetFigFont{12}{14.4}{\familydefault}{\mddefault}{\updefault}{\color[rgb]{0,0,0}$\le[\begin{array}{cc} 1 & w_{_L}(z) \\ 0 & 1\end{array}\ri]$}%
}}}}
\put(7615,472){\makebox(0,0)[lb]{\smash{{\SetFigFont{12}{14.4}{\familydefault}{\mddefault}{\updefault}{\color[rgb]{0,0,0}$\1$}%
}}}}
\put(3026,-661){\makebox(0,0)[lb]{\smash{{\SetFigFont{9}{10.8}{\familydefault}{\mddefault}{\updefault}{\color[rgb]{0,0,0}$\beta_j$}%
}}}}
\put(3133,-1454){\makebox(0,0)[lb]{\smash{{\SetFigFont{9}{10.8}{\familydefault}{\mddefault}{\updefault}{\color[rgb]{0,0,0}$|z|=1$}%
}}}}
\end{picture}%

%% file: TopDumbbell.pdf_t
\begin{picture}(0,0)%
\includegraphics{TopDumbbell.pdf}%
\end{picture}%
\setlength{\unitlength}{3947sp}%
\begingroup\makeatletter\ifx\SetFigFont\undefined%
\gdef\SetFigFont#1#2#3#4#5{%
  \reset@font\fontsize{#1}{#2pt}%
  \fontfamily{#3}\fontseries{#4}\fontshape{#5}%
  \selectfont}%
\fi\endgroup%
\begin{picture}(5350,3358)(3341,-2366)
\put(3593,-503){\makebox(0,0)[lb]{\smash{{\SetFigFont{12}{14.4}{\familydefault}{\mddefault}{\updefault}{\color[rgb]{0,0,0}$\beta_{j+1}$}%
}}}}
\put(7298,-2288){\makebox(0,0)[lb]{\smash{{\SetFigFont{12}{14.4}{\familydefault}{\mddefault}{\updefault}{\color[rgb]{0,0,0}$\beta_j$}%
}}}}
\end{picture}%